\documentclass[journal]{IEEEtran}
\usepackage{cite}
\usepackage{framed}

\ifCLASSINFOpdf
  \usepackage[pdftex]{graphicx}
  \graphicspath{{./images/}, {./photos/}}
  \DeclareGraphicsExtensions{.pdf,.jpeg,.png}
\else
\fi
\usepackage[cmex10]{amsmath}
\begin{document}
\title{Keypoint Encoding for Improved Feature Extraction from Compressed Video at Low Bitrates}

\author{Jianshu~Chao,~\IEEEmembership{Student~Member,~IEEE,}
        Eckehard~Steinbach,~\IEEEmembership{Fellow,~IEEE}% <-this % stops a space
        
\thanks{J. Chao and E. Steinbach are with the Chair of Media Technology, Technische Universit\"at M\"unchen, 80333 Munich, Germany. E-mails: \{jianshu.chao, eckehard.steinbach\}@tum.de.}
\thanks{Manuscript received April X, XXXX; revised December X, XXXX.}}


\markboth{Technical report}%
{Shell \MakeLowercase{\textit{et al.}}: Bare Demo of IEEEtran.cls for Journals}



\maketitle

\begin{abstract}
In many mobile visual analysis applications, compressed video is transmitted over a communication network and analyzed by a server. Typical processing steps performed at the server include keypoint detection, descriptor calculation, and feature matching. Video compression has been shown to have an adverse effect on feature-matching performance. The negative impact of compression can be reduced by using the keypoints extracted from the uncompressed video to calculate descriptors from the compressed video. Based on this observation, we propose to provide these keypoints to the server as side information and to extract only the descriptors from the compressed video. First, we introduce four different frame types for keypoint encoding to address different types of changes in video content. These frame types represent a new scene, the same scene, a slowly changing scene, or a rapidly moving scene and are determined by comparing features between successive video frames. Then, we propose Intra, Skip and Inter modes of encoding the keypoints for different frame types. For example, keypoints for new scenes are encoded using the Intra mode, and keypoints for unchanged scenes are skipped. As a result, the bitrate of the side information related to keypoint encoding is significantly reduced. Finally, we present pairwise matching and image retrieval experiments conducted to evaluate the performance of the proposed approach using the Stanford mobile augmented reality dataset and 720p format videos. The results show that the proposed approach offers significantly improved feature matching and image retrieval performance at a given bitrate. 

\end{abstract}


% \fi
\IEEEpeerreviewmaketitle

\section{Introduction}
\IEEEPARstart{T}HE extraction of features from images or videos is a fundamental component of many computer vision algorithms. Ideally, the feature extraction process identifies features that are shift-invariant, scale-invariant, rotation-invariant, illumination-invariant, etc. The extracted features are frequently compared with features in a database to identify correspondences. Typically, in the feature extraction process, keypoints (also called interest points or salient points) are detected first, and then, local descriptors (also called feature vectors) are calculated from the image patches located around these keypoints. With the increasing ubiquity of camera-equipped mobile devices and high-speed wireless communication networks, novel applications such as mobile visual search are emerging. In such applications, most of the feature-related processing is typically performed at a server. Note that feature extraction can be performed by either the client or the server. For client-side feature extraction, compressed features are uploaded to the server. The compression of the features is expected to create only small distortions compared with the uncompressed features. For server-side feature extraction, the images/videos are compressed and transmitted to the server. In this case, it is important to minimize the impact of the compression on the feature extraction performed at the server. Ideally, the features extracted from a compressed image or video should be identical, or at least very similar, to the features extracted from the uncompressed version. Both client- and server-based feature extraction approaches fall into the emerging area of feature-related compression approaches. 

Previous studies of feature-related compression can be categorized into three classes. The first class involves direct compression of the features. For example, several studies~\cite{Chandrasekhar:SurveySIFT,  Baroffio:Visual, Makar:Interframe14} have proposed methods of encoding scale-invariant feature transform (SIFT) features~\cite{Lowe:SIFT} extracted from images and video sequences. Other studies~\cite{Redondi:Compress, Redondi:Rate, Baroffio:CodingBinary} have proposed the encoding of binary features (e.g., BRISK features~\cite{Leutenegger:BRISK}), and algorithms of this type are named the \textit{analyze-then-compress (ATC)} paradigm.
%%%Editor - Please ensure that the intended meaning has been maintained in the
%%%above edit.
Along the same lines, the compact descriptors for visual search (CDVS)~\cite{MPEG:CDVS, Duan:DCC2015, Duan:CDVS} standard aims to standardize technologies for feature encoding at low bitrates. In the second class, canonical image patches are compressed and then transmitted or stored for further processing~\cite{Makar:patches, Makar:Quantization, Makar:Interframe, Makar:Interframe14}. Third is the standard image-compression-based architecture, which the authors of~\cite{Redondi:Compress} have dubbed the \textit{compress-then-analyze (CTA)} paradigm.
%%%Editor - Please ensure that the intended meaning has been maintained in the
%%%above edit.
Furthermore, a few approaches involve modifying standard image/video compression algorithms such that the features extracted from the compressed images/videos are as similar as possible to the features extracted from the uncompressed images/videos. These approaches (also belonging to the third class) are referred to as \textit{feature-preserving image and video compression}~\cite{Chao:Image, Chao:Video, Chao:Table2013, Chao:keypointImage}. To this end, the authors of~\cite{Duan:table2012, Chao:Table2013} optimize the JPEG quantization table, whereas in~\cite{Chao:Image, Chao:Video}, the rate allocation strategy is modified to preserve the most important features. Recently, \cite{Chao:keypointImage} has proposed encoding the SIFT keypoints and transmitting them as side information along with the compressed image. Similarly, \cite{Baroffio:ICIP2015} proposed transmitting the encoded BRISK keypoint locations, scales, and differential BRISK descriptors along with the image to a server. In both approaches, the keypoints are sent as side information for improved feature extraction from compressed images. Compared with the first two classes, the advantages of feature-preserving image/video compression are that the decoded images/videos can also be viewed and stored for future use and that other types of features can later be extracted. For solutions providing standard-compatible images/videos, such as~\cite{Duan:table2012, Chao:Table2013, Chao:Image, Chao:Video, Chao:keypointImage, Baroffio:ICIP2015}, standard decoders can be used to decode the images/videos. In all studies mentioned above, the objective was to achieve low-bitrate data transmission for applications such as mobile visual search, video surveillance, and visual localization.

\begin{figure*}[!htb]
%\vspace{-0.1in}
\centering
\includegraphics[width=0.8 \textwidth]{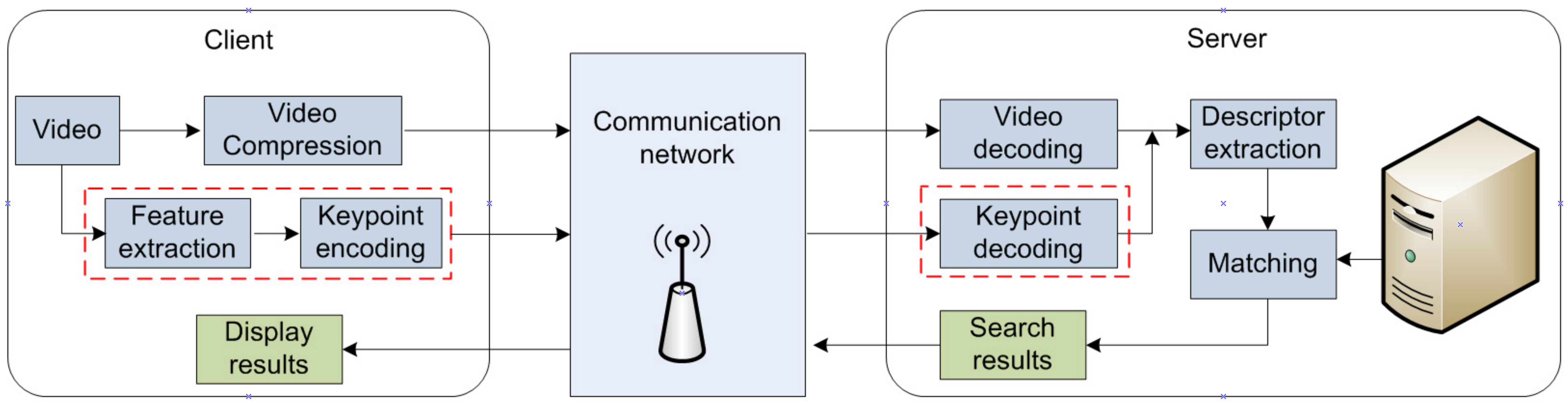} 
%\vspace{-0.1in}
\caption{Overview of the proposed approach. The keypoints extracted from the video content at the client are encoded and transmitted as side information along with the compressed video to the server.}
\vspace{-0.1in}
\label{fig:idea_architecture}
\end{figure*}

Few studies have addressed the compression of features extracted from videos. In~\cite{Baroffio:Visual, Baroffio:CodingBinary}, the authors proposed intra- and inter-frame coding modes for SIFT descriptors and binary local descriptors, respectively, to encode the descriptors extracted from a video sequence. The authors of~\cite{Makar:Interframe} proposed inter-frame predictive coding techniques for image patches and keypoint locations, and they subsequently proposed an \textit{inter-descriptor coding} scheme~\cite{Makar:Interframe14} to encode the descriptors extracted from such patches. In these approaches, descriptors/patches are extracted from videos and compressed for transmission; however, the videos themselves are not stored or sent to a server. By contrast, in this work, we transmit compressed videos via a communication network to a server to perform feature extraction. The advantages previously discussed with respect to images also apply to videos. Based on our previous study regarding images~\cite{Chao:keypointImage}, we propose in this work a predictive keypoint encoding approach to encode the original keypoints extracted from uncompressed videos. In the proposed approach, the compressed keypoints are sent as side information along with the compressed videos at low bitrates. At the server, we use the decoded keypoints (locations, scales, and orientations) to extract feature descriptors from the videos. The proposed approach is illustrated in Fig.~\ref{fig:idea_architecture}. The proposed framework is fundamentally different from those reported in other studies~\cite{Baroffio:Visual, Makar:Interframe14}, which encode and transmit the descriptors. By contrast, in the proposed approach, only the keypoints are encoded and transmitted along with the compressed video. To evaluate the proposed approach, we conduct experiments using the widely used H.264/AVC and H.265/HEVC standards for video encoding. The results are presented as plots that show the number of successfully matched features versus the bitrate. In addition, we show the percentages of images that are successfully retrieved using various approaches via a content-based image retrieval engine. 

%Our approach achieves better matching performance than the patch encoding approaches in~\cite{Makar:Interframe} and~\cite{Makar:Interframe14}, while the decoded videos are still available for human observer.

The remainder of this paper is organized as follows. In Section II, the impact of video compression on feature quality is examined. The feature-preservation performance of H.264/AVC and H.265/HEVC as a function of bitrate is presented, and the motivation for the proposed approach is explained in greater detail. In Section III, we introduce four different types of frames for keypoint encoding based on the changes in the video content and propose keypoint encoding approaches for each of these different frame types. Section IV presents the details of the proposed keypoint prediction framework. In Section V, we detail the Intra, Skip and Inter modes of keypoint encoding and transmission, which allow us to significantly reduce the bitrate of the side information. In Section VI, we present pairwise matching and image retrieval experiments. Our results show that the proposed approach offers substantially improved performance compared with that of standard H.265/HEVC-encoded videos. Conclusions are presented in Section VII. 

\section{Impact of video compression on feature quality}
\label{sec:comparison}
In this section, we investigate the impact of video compression on the features extracted from compressed videos. For this purpose, we first extract SIFT features from videos encoded using H.264/AVC and H.264/HEVC and compare these features with the features extracted from a set of uncompressed reference images. Then, we plot the number of matches as a function of bitrate and compare the results for the different video compression standards.

\subsection{Experimental design} \label{exprimentalSettings}
For the matching performance evaluation, we choose the Stanford mobile augmented reality (MAR) dataset~\cite{Makar:Interframe}, which comprises 23 videos (each containing a single static object) and 23 corresponding reference images. Each video consists of 100 frames (30 fps) at a resolution of 640 $\times$ 480. Similar to~\cite{Makar:Interframe}, we use eight video sequences (\textit{OpenCV}, \textit{Wang Book}, \textit{Barry White}, \textit{Janet Jackson}, \textit{Monsters Inc.}, \textit{Titanic}, \textit{Glade}, and \textit{Polish}) for pairwise feature-matching evaluations. Furthermore, we use SIFT features and the Vlfeat~\cite{Vedaldi:vlfeat} SIFT implementation. Similar to~\cite{Makar:Interframe}, the top 200 features are selected for each frame in accordance with the CDVS Test Model. The nearest-neighbor distance ratio (NNDR) is used to evaluate matching descriptors. For a query descriptor $D$ extracted from a compressed test frame, the nearest descriptor $D_{a}$ and the second nearest descriptor $D_{b}$ from the reference image are found, and thresholding is applied to their distance ratio. The query descriptor $D$ and the nearest descriptor $D_{a}$ are considered to match if $||D-D_{a}||/||D-D_{b}|| < t$. We calculate the Euclidean distances between the descriptors and set $t$ to 0.8~\cite{Lowe:SIFT}. Then, we use random sample consensus (RANSAC)~\cite{Fischler:RANSAC} to remove incorrectly matched features, assuming an affine transformation between the reference image and the test frame. 
%
%For retrieval performance evaluation, we use video sequences that show multiple objects in the Stanford MAR dataset. Each \textit{Multiple Objects} video has 200 frames and contains three different objects of interest. The first two \textit{Multiple Objects} videos are used in our retrieval experiments. Excluding the fast spatial matching component, we use a previously proposed image retrieval system~\cite{Philbin:AKM}. We use the MIRFLICKR-25000~\cite{Huiskes:MIR} database and the 23 reference images from the Stanford MAR dataset as the training dataset. Similar to~\cite{Makar:Interframe}, we extract up to 300 SIFT descriptors for each image in the database and train one million visual words (VW) from these descriptors. For test frames, we extract 200 SIFT features and feed them to the retrieval engine. After obtaining a shortlist of candidate matched images from the retrieval system, we run RANSAC for the top 100 images in the shortlist to reorder the retrieved images for improved precision. 

\subsection{Feature-matching performance for different video compression standards}
\label{subsec:matchesStandards}
The eight test video sequences are encoded using the JM reference software~\cite{HHI:JM} and the HEVC Test Model~\cite{HHI:HEVC}. To be able to compare our results with the patch-encoding approaches presented in~\cite{Makar:Interframe, Makar:Interframe14}, we use the same parameters as in~\cite{Makar:Interframe, Makar:Interframe14} for H.264/AVC. Additionally, we use more QP values to produce high-quality videos. The settings are as follows: the IPPP$\cdots$ structure, IntraPeriod = 50 frames, QP$_{Pframes}$ = \{26, 30, 34, 38, 42, 46, 50\}, and QP$_{Iframes}$ = QP$_{Pframes}$ -3. The parameters used for H.265/HEVC encoding are the same as those in the example provided in the HEVC version 16.0 manual \cite{HHI:HEVC}. The GOP structure is IBBBPBBBP$\cdots$, and QP = \{22, 26, 30, 34, 38, 42, 46, 50\}.

\begin{figure}[!htb]
\vspace{-0.1in}
\centering
\includegraphics[width=0.48 \textwidth]{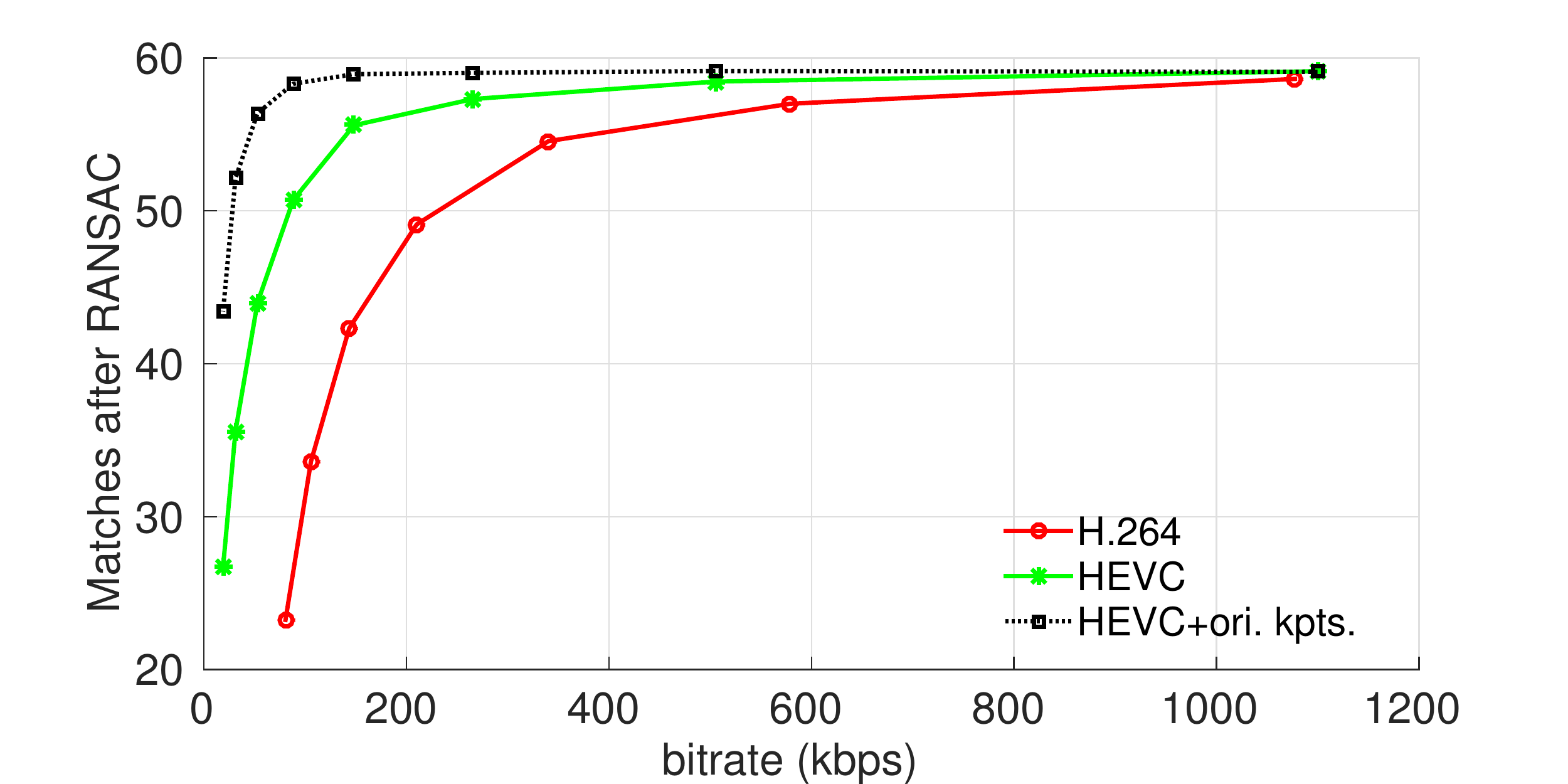} 
%\vspace{-0.1in}
\caption{SIFT-feature-matching performance for different video compression schemes. The dotted line shows the results obtained for descriptors calculated from the H.265/HEVC encoded videos using the original keypoints extracted from the uncompressed video frames.}
\vspace{-0.1in}
\label{fig:results_video_ideal}
\end{figure}

In the following experiments, SIFT features are extracted from the compressed frames and compared with the SIFT features extracted from the corresponding uncompressed reference images. The number of matching features after RANSAC is applied and the bitrates for the encoded videos are averaged over all test frames. The solid red and green curves in Fig.~\ref{fig:results_video_ideal} represent the feature-matching performance for H.264/AVC- and H.265/HEVC-encoded videos as a function of bitrate. The H.265/HEVC-encoded videos exhibit much better performance than the videos encoded with H.264/AVC in terms of the number of feature matches at a given bitrate.

\subsection{Sensitivity of keypoints and descriptors}
Given an input video frame $I$, the detected keypoints can be expressed as
\begin{equation}
k_i=\{l_i, \sigma_i, \theta_i\}
\end{equation}
where $l_i$ is the location of keypoint $i$ in the frame, $\sigma_i$ is the scale, and $\theta_i$ is the orientation of the keypoint. The descriptors obtained using the keypoints detected in frame $I$ are expressed as follows:
\begin{equation} \label{eq:k_i_d_i}
d_i = \Psi(k_i | I)
\end{equation}
where $\Psi$ represents the descriptor extraction operation on $I$ using the keypoints $k_i$. Similarly, the descriptors extracted from the compressed frame $\overline{I}$ are expressed as
\begin{equation} \label{eq:overline_k_i_I}
\overline{d_i} = \Psi(\overline{k_i} | \overline{I})
\end{equation}
Here, the $\overline{k_i}$ are the keypoints detected in $\overline{I}$. Similar to our previous study on keypoint encoding for improved feature extraction from compressed images~\cite{Chao:keypointImage}, we perform a simple experiment to demonstrate that the results of keypoint detection can be easily affected by video compression artifacts and that descriptors are more robust. To this end, we calculate the descriptors from the compressed video frames using the keypoints extracted from the original (uncompressed) frames. This procedure is expressed as follows: 
\begin{equation} \label{idealD}
d_i^{\prime} = \Psi({k_i} | \overline{I})
\end{equation}

This means that the features $d_i^{\prime}$ have exactly the same keypoints $k_i$ as in the uncompressed case. This allows us to ignore the possibility of inaccurate keypoint detection from the compressed frame and evaluate exclusively the descriptor robustness in the presence of compression artifacts. Because the original keypoints $k_i$ cannot be obtained from the compressed frame $\overline{I}$, we present the matching results obtained based on H.265/HEVC encoded videos as a dotted line (upper bound) in Fig.~\ref{fig:results_video_ideal}. For videos of higher quality, the gap between the \textit{HEVC} approach and the \textit{HEVC+ori. kpts.} approach in Fig.~\ref{fig:results_video_ideal} becomes increasingly smaller. However, high-quality videos also require a much higher bitrate and can be transmitted only if the necessary network resources are available. To avoid excessive bandwidth requirements for applications such as mobile visual search or video surveillance, we must compress the data to a low bitrate for transmission. This is the motivation for the current research on feature compression algorithms in the literature as well as the MPEG CDVS standard. Thus, our goal is not to provide high-quality video for human observers. Instead, we are interested in videos encoded at low bitrates for communication networks of limited capacity. The results indicate that if the original keypoints are preserved, then the feature-matching performance can be improved, especially for strong compression (i.e., low-bitrate encoding). This observation motivates us to encode the keypoints and send them as side information along with such videos compressed to low bitrates. In the following experiments, we will use QP = \{38, 42, 46, 50\} for H.264/AVC and H.265/HEVC encoding. 

% The results indicate that if the original keypoints are preserved, the descriptor vectors can be well preserved even for strong compression (i.e., low bitrate encoding).

\section{Core ideas}
In our previous study~\cite{Chao:keypointImage}, we presented a keypoint encoding approach for still images. Applying this approach directly to individual frames in a video sequence would significantly increase the bitrate, as will be discussed in Section~\ref{sec:keypointencoding}. To address this issue, similar to the conventional inter-frame prediction scheme in video coding, we propose several keypoint prediction approaches that significantly reduce the number of keypoints to be encoded and thus the bitrate required for the side information. 

%%%%%%%%%%%%%%%%%%%%%%%%%%%%%%%%%%%% move to a suitable location
\begin{figure*}[!htb]
\centering
\includegraphics[width=0.78  \textwidth]{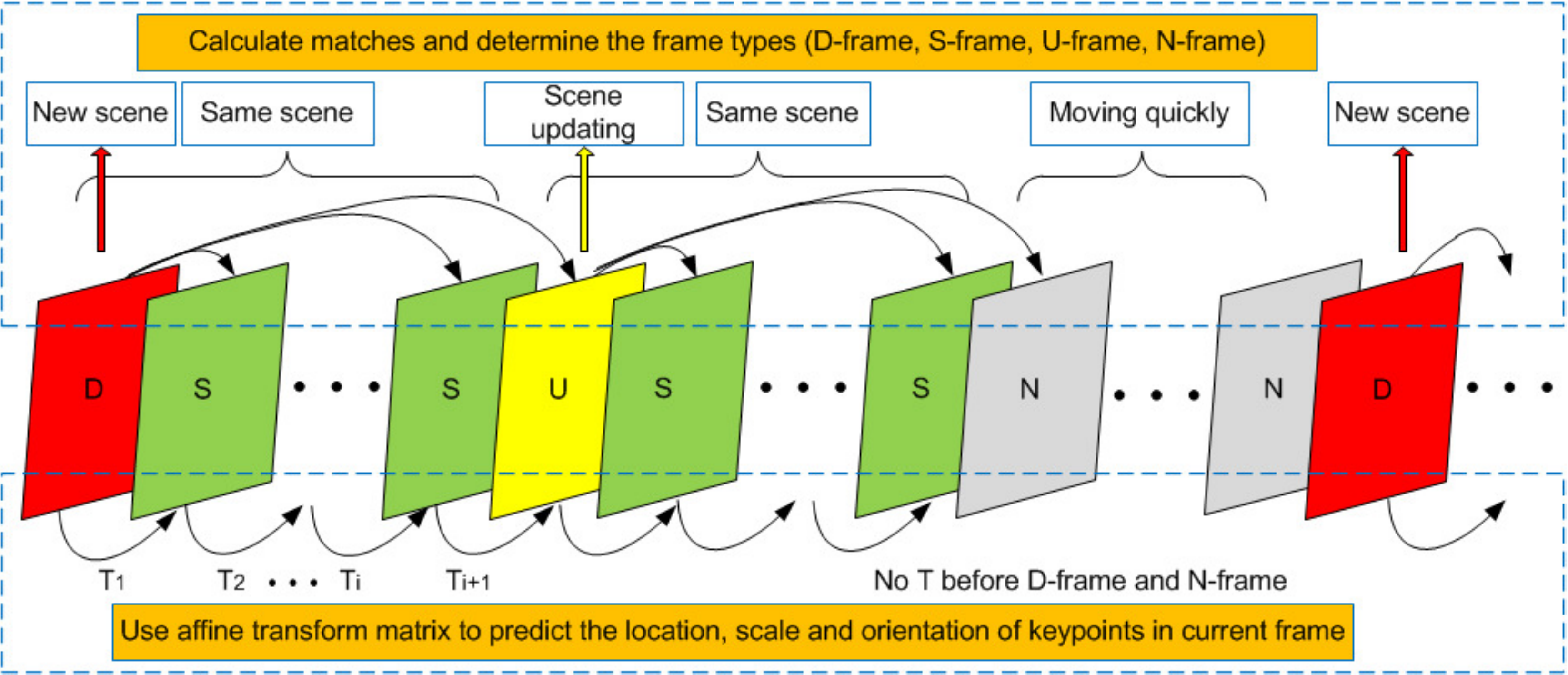} 
%\vspace{-0.1in}
\caption{Examples of the four types of frames for keypoint encoding.}
%\vspace{-0.1in}
\label{fig:frameTypes}
\end{figure*}  

%%%%%%%%%%%%%%%%%%% move to a suitable position
\begin{figure*}[!htb]
\centering
\includegraphics[width=0.99  \textwidth]{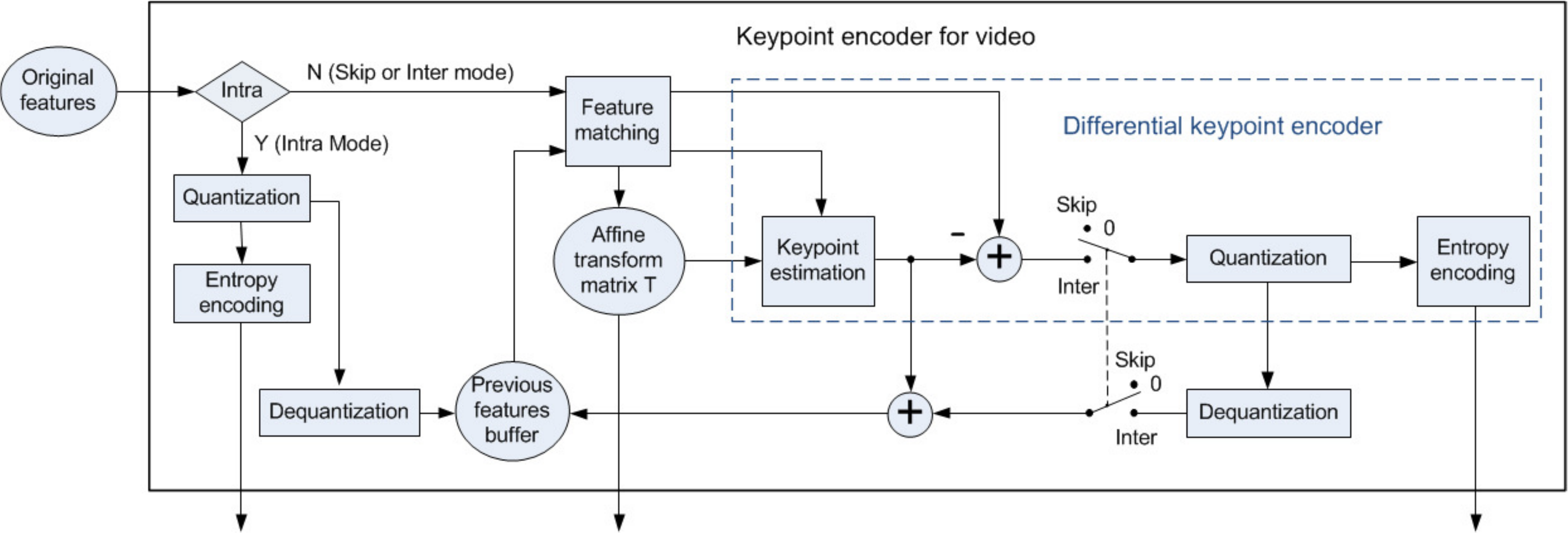} 
%\vspace{-0.1in}
\caption{Keypoint encoding for video.} 
\vspace{-0.1in}
\label{fig:KeypointEncoderVideo}
\end{figure*}

\subsection{Frame types for keypoint encoding}
The locations, scales, and orientations of keypoints detected in consecutive frames are related. If the keypoints are detected independently for each frame, then some of the keypoints may disappear or reappear across the frames as a result of the feature detection process. However, a keypoint that has disappeared from the previous frame may still yield a useful descriptor for the current frame. To address this issue, the authors of \cite{Makar:Interframe} proposed a \textit{temporally coherent keypoint detector} in which the detected patches are propagated to consecutive video frames. Unlike their approach, we extract keypoints and use them to predict the keypoints (locations, scales, and orientations) in consecutive video frames because, in our case, the video itself is also available. Here, we introduce four different types of frames for keypoint encoding. These frame types are illustrated in Fig.~\ref{fig:frameTypes}. Two of the four frame types are similar to those proposed in~\cite{Makar:Interframe}: \textit{Detection frames} (\textit{D-frames}) and \textit{Skip frames} (\textit{S-frames}). For D-frames, the keypoints are extracted using a conventional feature detection process. For S-frames, all keypoints are estimated using the keypoints from the previous frame, and the direct encoding of the keypoints is skipped. Therefore, S-frames enable a significant reduction in the amount of side information, as discussed in the following section. After a certain number of frames, it might no longer be possible to estimate the keypoints of the current frame using the previously identified keypoints because, for example, the object of interest is leaving the field of view or the estimated keypoints do not yield effective descriptors calculated from the current video frame. To address this case, we add a third type of video frame, termed an \textit{Update frame} (\textit{U-frame}), to update the keypoints. In contrast to D-frames, U-frames combine both conventionally detected keypoints and forward-estimated keypoints because a few of the estimated keypoints are still sufficient for calculating the descriptors. However, we find that if the scene is moving quickly, then even U-frames will generate a large number of bits. To resolve this issue, we add a fourth type of frame, called a \textit{Null frame} (\textit{N-frame}), for which keypoint encoding and transmission are switched off. For N-frames, no side information is transmitted, and the features are extracted exclusively from the compressed frames. The method of determining the frame types and the keypoint encoding processes for the different frame types are explained in the following sections.

\subsection{Keypoint encoder}
As previously described, we predict the keypoints for S- and U-frames from the keypoints in the previous frame, as illustrated in Fig.~\ref{fig:KeypointEncoderVideo}. Note that the keypoint decoder follows the reverse process. Below, we define the terminology that will be used throughout this paper for clarity. 

\noindent $\bullet$ \textit{Estimate} means to estimate the locations, scales and orientations of the keypoints in the current frame using keypoints from previous frames. 

\noindent $\bullet$ \textit{Update} means to estimate the keypoints and also use differential keypoint encoding to update the keypoints. 

\noindent $\bullet$ \textit{Predict} means to either estimate or update the keypoints. Note that although estimate and predict are synonyms, they are used for different purposes in this paper. 

\noindent $\bullet$ In \textit{Intra mode}, the keypoints are extracted by conventional means from the current frame without reference to keypoints from previous frames and are encoded using a process similar to that presented in ~\cite{Chao:keypointImage}.

\noindent $\bullet$ In \textit{Skip mode}, the keypoints are estimated and then stored in the buffer. Thus, differential keypoint encoding and transmission are skipped. 

\noindent $\bullet$ In \textit{Inter mode}, the keypoints are estimated and the differential encoder is used to update their locations, scales and orientations. Afterward, they are stored in the buffer. 

Note that only the Intra mode is used to encode the keypoints for D-frames; only the Skip mode is used for S-frames; all three modes (the Intra, Skip, and Inter modes) are used for U-frames; and no keypoint encoding is performed for N-frames. We will explain the components of the diagram in Fig.~\ref{fig:KeypointEncoderVideo} in detail in the following sections.

\section{Keypoint prediction}
\label{sec:keypointPrediction}
\begin{figure}[!htb]
\centering
\includegraphics[width=0.48 \textwidth]{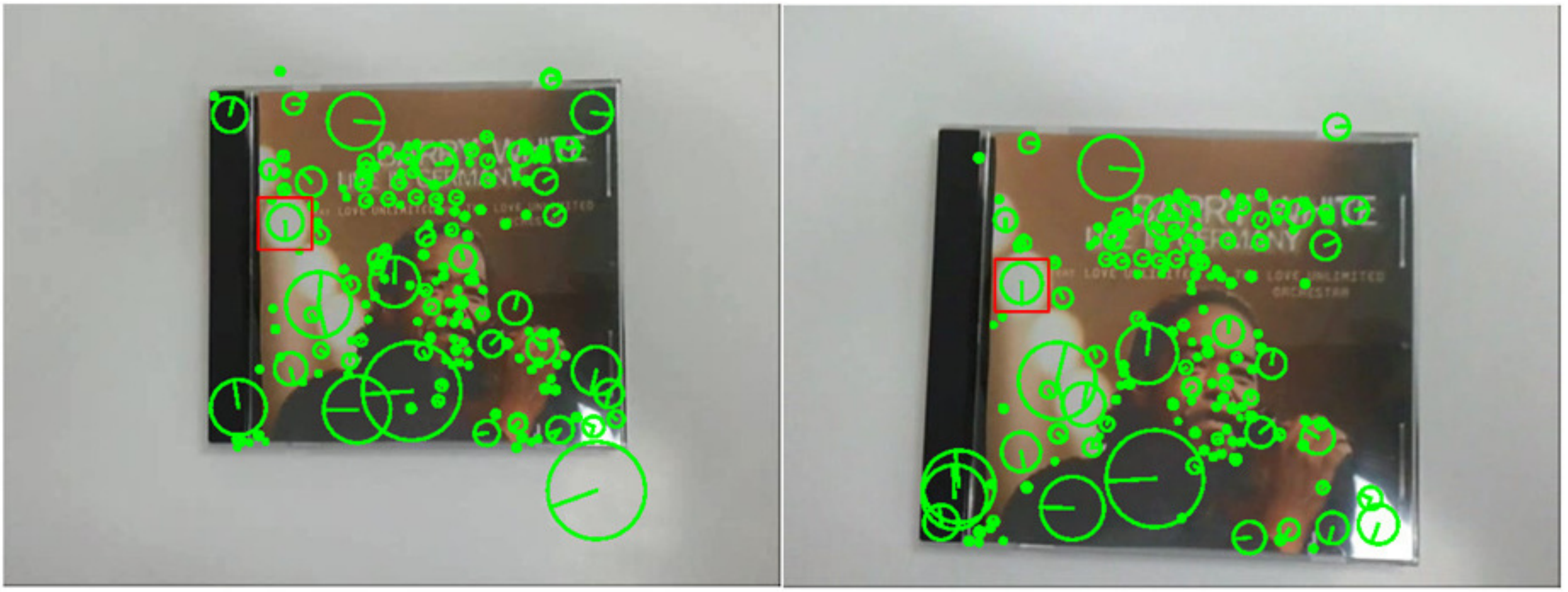} 
%\vspace{-0.1in}
\caption{Keypoints from video frame 1 and video frame 20 in the example video \textit{Barry White}.} 
\vspace{-0.1in}
\label{fig:Keypoints_twoFrames}
\end{figure}

When the camera or the object in a video is moving, the locations, scales and orientations of the detected keypoints are also gradually changing. Fig.~\ref{fig:Keypoints_twoFrames} shows the keypoints detected in frame 1 and frame 20 of the video sequence \textit{Barry White}. We can see that the keypoints are still closely related. The red squares indicate a pair of related keypoints. However, the location, scale and orientation of the keypoint have changed; thus, we must predict the keypoints for the current frame using the previous keypoints. 

\subsection{Feature matching and affine transformation}
As shown in Fig.~\ref{fig:frameTypes}, we determine the frame types for keypoint encoding by matching the features of the current frame to the features of the previous frame. To simplify the discussion, let us first consider only D-frames and S-frames to describe our proposed framework. As introduced in the previous section, the keypoints in D-frames are detected without reference to previous frames (Intra mode). We estimate the keypoints for the subsequent S-frames using all keypoints from the previous frame. The keypoint estimation process for S-frames is represented by \textit{Skip} in Fig.~\ref{fig:KeypointEncoderVideo}. The number of frames between consecutive D-frames is the \textit{detection interval $\Delta$}, which is set to a fixed number (e.g., 5 or 20) in our first experiment. 

Two video frames have a relationship that can be locally described by a geometric transformation, e.g., an affine transformation or a perspective transformation. In our experiments, we assume that the relationship between two consecutive frames can be described by an affine transformation and transmit the corresponding affine transform matrix. The locations, scales, and orientations of the keypoints in the current frame can then be estimated using this transform and the keypoints from the previous frame. The features from the previous frame are stored in the \textit{Previous features buffer}. The block \textit{Feature matching} is used to calculate the affine transform matrix. In the literature, several affine transform matrices were determined in~\cite{Wiegand:affine} for subregions of a video frame with the goal of improving the conventional rate-distortion performance in video coding. We can also use several affine transform matrices in the case that the video contains multiple objects of interest that are moving independently. In the Stanford MAR dataset, all of the test videos contain objects of interest that are moving in the same direction. Thus, similar to~\cite{Makar:Interframe14}, we use only one common transform matrix in this paper. However, multiple matrices can be used for complex video scenes to improve performance. Next, we will explain the blocks labeled \textit{Affine transform matrix T} and \textit{Keypoint estimation} in Fig.~\ref{fig:KeypointEncoderVideo} in detail.

\subsection{Location, scale, and orientation estimation}
\label{sec:keypointAdaptation}
After obtaining the affine transform matrix, we estimate the keypoints in the current frame using the keypoints from the previous frame. The estimated location can be easily calculated as follows:
 
% If the affine transform cannot be obtained using the keypoints from two consecutive frames, it means that a new scene has begun and we choose a D-frame, which is discussed in Section~\ref{subsec:adaptive}. 

\begin{equation} \label{eq:affineTransform}
\begin{aligned}
\begin{bmatrix} \hat x  \\ \hat y \\ 1  \end{bmatrix} & = T \cdot \begin{bmatrix} x \\y \\ 1 \end{bmatrix} \\
& = \begin{bmatrix}a & b & t_x \\ c & d & t_y \\ 0 & 0 & 1 \end{bmatrix}\begin{bmatrix} x \\y \\ 1 \end{bmatrix} \\
& = \begin{bmatrix}1 & 0 & t_x \\ 0 & 1 & t_y \\ 0 & 0 & 1 \end{bmatrix}  \begin{bmatrix}a & b & 0 \\ c & d & 0 \\ 0 & 0 & 1 \end{bmatrix}  \begin{bmatrix} x \\y \\ 1 \end{bmatrix}
\end{aligned}
\end{equation}
where $(x, y)$ is the keypoint location in frame $f-1$ and $(\hat x, \hat y)$ is the estimated location in frame $f$. \eqref{eq:affineTransform} represents the affine transformation in homogeneous coordinates. Here, $T$ is the affine transform matrix, and we first decompose the transform matrix into two component matrices as shown in~\eqref{eq:affineTransform}. The second matrix can be further decomposed as follows:
\begin{equation} \label{eq:affineA}
\begin{aligned}
A & = \begin{bmatrix}a & b \\c & d \end{bmatrix} \\
& = \begin{bmatrix}r_1 & 0 \\ 0 & r_2 \end{bmatrix} \begin{bmatrix}1 & 0 \\q & 1 \end{bmatrix} \begin{bmatrix}cos(\phi) & sin(\phi) \\ -sin(\phi) & cos(\phi) \end{bmatrix}
\end{aligned}
\end{equation}
where the matrix on the left represents the scaling, the middle matrix represents the shearing, and the matrix on the right represents the rotation. Note that this decomposition is not unique. Here, $r_1$ and $r_2$ are the scaling factors in two directions, $q$ is the shearing factor, and $\phi$ is the clockwise rotation angle. By solving the four linear equations represented by~\eqref{eq:affineA}, we obtain the following:
\begin{equation} \label{eq:decomposition}
\begin{aligned}
r_1 & = \sqrt{a^2+b^2} \\
r_2 & =  \frac{ad-bc}{\sqrt{a^2+b^2}} \\  
q  & = \frac{ac+bd}{ad-bc} \\
\phi & = atan(b, a) 
\end{aligned}
\end{equation}

\begin{figure}[!htb]
\centering
\vspace{-0.1in}
\includegraphics[width=0.35 \textwidth]{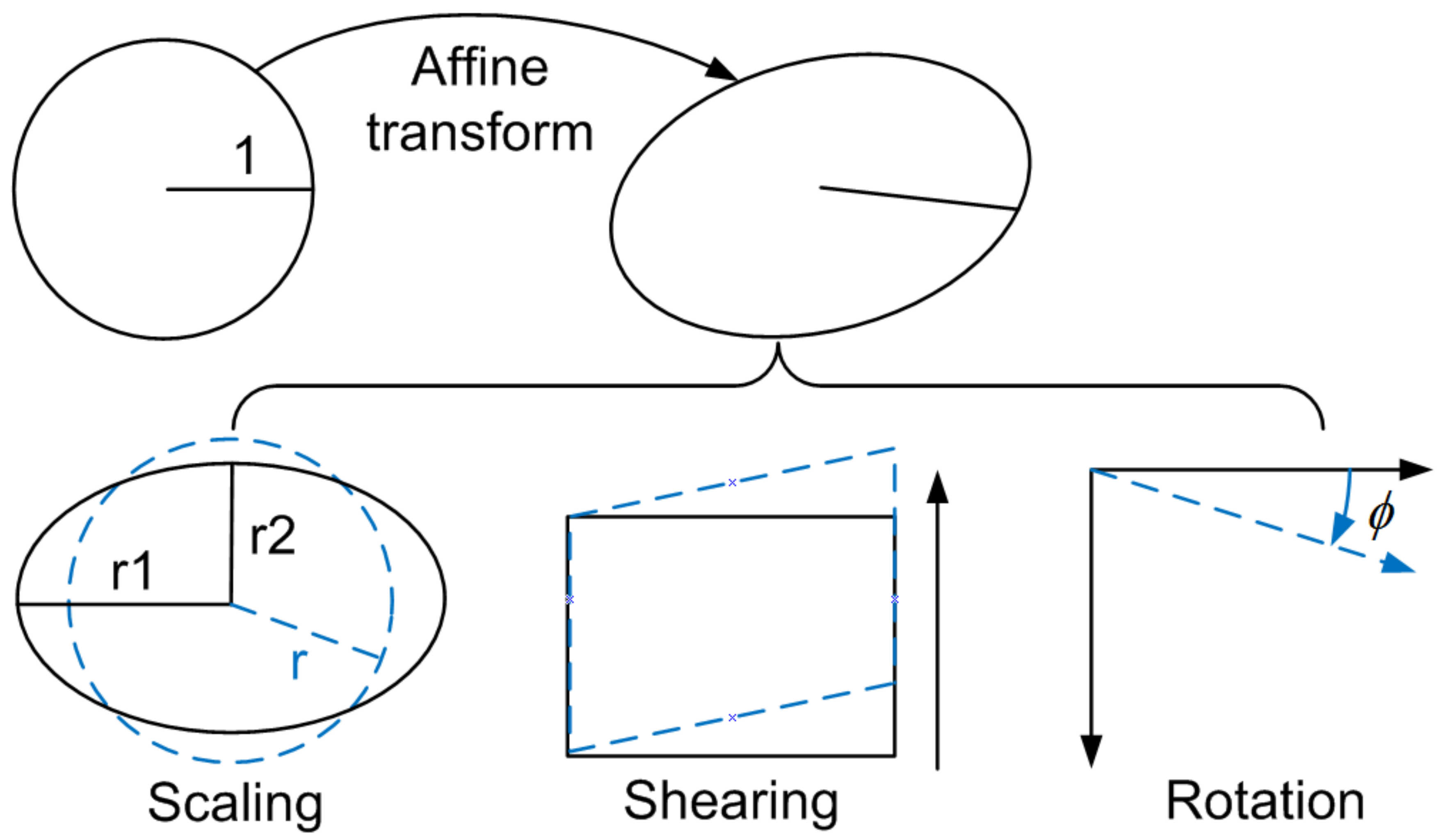} 
\caption{Scaling, shearing, and rotation transformations derived from matrix A.}
\label{fig:decomposition}
\end{figure}

These formulae are graphically illustrated in Fig.~\ref{fig:decomposition}. After the affine transformation, the circular keypoint in the first frame has taken on an elliptic shape in the second frame. Its area is determined by the scaling factors $r_1$ and $r_2$ and the shearing factor $q$ in~\eqref{eq:affineA}. From Fig.~\ref{fig:decomposition}, we can see that the shearing and rotation transformations do not change the area, whereas $r_1$ and $r_2$ do affect the area. Because the keypoint of a SIFT feature has a circular shape, we assume that the elliptical area of an estimated keypoint should be the same as the area of the corresponding keypoint (the dashed blue circle with radius $r$) in the current frame that would be detected using the conventional detection process. From the formulae for calculating the areas of a circle and an ellipse, the following scaling factor is obtained:
\begin{equation} \label{eq:scaling}
\begin{aligned}
\pi r^2 & = \pi r_1 r_2 \\
r & = \sqrt{r_1 r_2} \\
s & = \dfrac{r}{1} = r
\end{aligned}
\end{equation}

\begin{figure}[!htb]
\centering
\vspace{-0.1in}
\includegraphics[width=0.48 \textwidth]{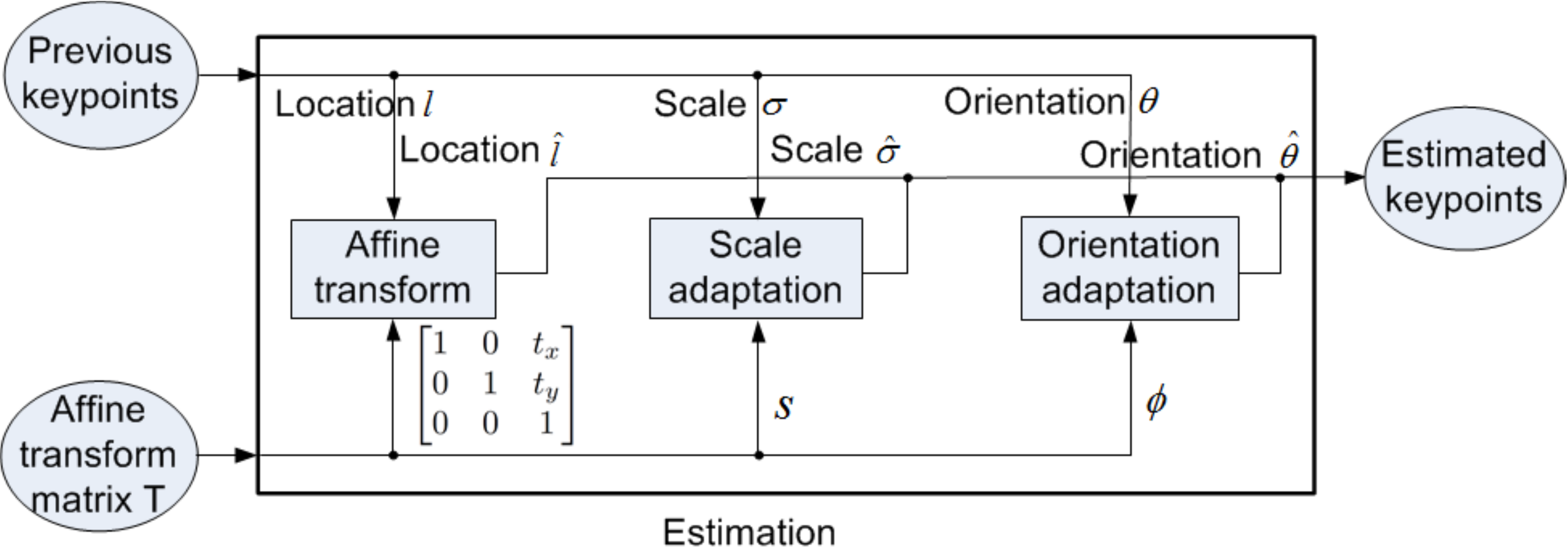} 
\caption{The location, scale and orientation estimation.} 
\label{fig:estimation}
\end{figure}

To summarize, the location $\hat{l}^{f}$ (a vector of $\hat{x}$ and $\hat{y}$) of a possible keypoint $\hat{k}^{f}$ in the current frame $f$ can be estimated by applying $T$ in~\eqref{eq:affineTransform} to the location $l^{f-1}$ (a vector of $x$ and $y$) of a keypoint $k^{f-1}$ in the previous frame $f-1$. This operation is expressed as follows:
\begin{equation} \label{eq:updateLocation}
\begin{bmatrix} \hat{l}^{f} \\ 1  \end{bmatrix} = T \cdot \begin{bmatrix} l^{f-1} \\ 1 \end{bmatrix} \\
\end{equation}
The scale of the keypoint is estimated by multiplying the scale of the original keypoint, $\sigma$, by the scaling factor $s$ as follows:
\begin{equation} \label{eq:updateScale}
\hat{\sigma}^{f} = s \cdot \sigma^{f-1} = \sqrt{r_1 r_2} \cdot \sigma^{f-1}
\end{equation}
Furthermore, the orientation is estimated by rotating the orientation $\theta^{f-1}$ as follows:
\begin{equation} \label{eq:updateOrientation}
\hat{\theta}^{f} = \theta^{f-1} - \phi  
\end{equation}
Thus, the block labeled \textit{Keypoint estimation} in Fig.~\ref{fig:KeypointEncoderVideo} can be detailed as shown in Fig.~\ref{fig:estimation}.

We perform an initial experiment to justify our method of estimating the location, scale, and orientation of a keypoint. In this experiment, we use the unquantized keypoints from the D-frames and the unquantized affine transform matrix $T$. The descriptors are calculated from the uncompressed video frames using the estimated keypoints in the S-frames to demonstrate the accuracy of the estimation. As described in Section~\ref{exprimentalSettings}, eight videos, each containing a single static object, are used; the SIFT descriptors from the video frames are compared with those extracted from the corresponding reference images, and the average numbers of matching features are recorded. 
 
\begin{table}[!htb]
%\vspace{-0.1in}
\caption{Evaluation of location, scale, and orientation estimation for keypoints}
\label{table:keypointAdaptation}
\centering 

% Preview source code for paragraph 2

\begin{tabular}{|c|c|c|}
\hline 
Adaptation & \multicolumn{2}{c|}{Avg. \# matches}\tabularnewline
\hline 
  & $\Delta$=5 & $\Delta$=20\tabularnewline

\hline 
\hline 
loc. only, \eqref{eq:updateLocation} & 58.33 & 49.20\tabularnewline
\hline 
loc. + sc., \eqref{eq:updateLocation} + \eqref{eq:updateScale} & 58.55 & 50.98\tabularnewline
\hline 
loc. + orient., \eqref{eq:updateLocation} + \eqref{eq:updateOrientation} & 58.96 & 57.64\tabularnewline
\hline 
loc. + sc. + orient., \eqref{eq:updateLocation} to~\eqref{eq:updateOrientation} & 59.19 & 59.13\tabularnewline
\hline 
Independent detection & \multicolumn{2}{c|}{59.13}\tabularnewline
\hline 
\end{tabular}

%\begin{tabular}{|c|c|}
%\hline 
%Adaptation & \multicolumn{1}{c|}{Avg. \# matches}\tabularnewline
%\hline 
%\hline 
%loc. only \eqref{eq:updateLocation} & 58.33\tabularnewline
%\hline 
%loc. + sc. \eqref{eq:updateLocation} +\eqref{eq:updateScale} & 58.55\tabularnewline
%\hline 
%loc. + orient.  \eqref{eq:updateLocation} + \eqref{eq:updateOrientation}   & 58.96\tabularnewline
%\hline 
%loc. + sc. + orient. \eqref{eq:updateLocation} to~\eqref{eq:updateOrientation}  & 59.19\tabularnewline
%\hline 
%Independent detection & 59.13\tabularnewline
%\hline 
%\end{tabular}

%\vspace{-0.1in}
\end{table}
We perform this experiment to verify the performance of the blocks labeled \textit{Affine transform}, \textit{Scale adaptation}, and \textit{Orientation adaptation} in the diagram shown in Fig.~\ref{fig:estimation}. In Table~\ref{table:keypointAdaptation}, \textit{loc. only} refers to the case in which only the keypoint locations are estimated using~\eqref{eq:updateLocation}, whereas the scales and orientations are simply estimated as the scales and orientations of the previous keypoints, i.e., $\hat{\sigma}^{f} = \sigma^{f-1}$ and $\hat{\theta}^{f} = \theta^{f-1}$. The notation \textit{loc. + sc.} refers to the case in which the scales are also modified using a scaling factor, as in~\eqref{eq:updateScale}. The notation \textit{loc. + sc. + orient.} indicates the case in which the locations, scales, and orientations are all appropriately transformed. In addition, the results of conventional keypoint detection and descriptor calculation for each separate frame are provided for comparison (\textit{Independent detection}). In this case, all frames are D-frames, and the keypoints are detected independently for each frame. Table~\ref{table:keypointAdaptation} shows that the matching performance is improved by applying~\eqref{eq:updateScale} and~\eqref{eq:updateOrientation}, indicating that our estimation method is effective. Note that for a detection interval of $\Delta$=5, \textit{loc. only} also achieves a high number of matches, and the improvement achieved for \textit{loc. + sc. + orient.} seems quite small because of the small detection interval and because the test videos each contain only a single object. However, when the detection interval is large ($\Delta$=20) or the video content is rapidly changing, \textit{loc. only} and \textit{loc. + sc.} can easily fail to produce correct descriptors. Accordingly, we can see that \textit{loc. + sc. + orient.} achieves much better performance in the third column of Table~\ref{table:keypointAdaptation}. In addition, two possible reasons that the keypoints estimated using \eqref{eq:updateLocation} to~\eqref{eq:updateOrientation} slightly outperform the independently detected keypoints are that the RANSAC computation yields inconsistent numbers of matching features and that keypoints may therefore disappear or reappear in consecutive frames as a result of the feature detection and feature selection processes. The estimated keypoints, which are absent in conventional feature detection, are still sufficient to extract effective descriptors. As a result, the average number of matching features obtained from the estimated keypoints could be slightly higher than that obtained using the independent detection method. In the following experiments, we use the estimated keypoints (location, scale, and orientation) to extract descriptors.

\subsection{Quantization of the affine transform matrix}
The affine transform matrix contains real-valued numbers; therefore, we must perform lossy encoding for this matrix. The affine transform parameters are quantized via scalar quantization in~\cite{Wiegand:affine}. The authors of~\cite{Makar:Interframe14} encode the affine transform matrix $T$ using differential keypoint location coding. They uniformly quantize the parameters $a$, $b$, $c$, $d$, $t_x$, and $t_y$ in~\eqref{eq:affineTransform} using 7 bits, resulting in 42 bits in total for the matrix $T$. By contrast, we propose to quantize the parameters $r_1$, $r_2$, $q$, and $\phi$ in~\eqref{eq:decomposition}. The associated keypoints in consecutive frames have similar scales and small differences in orientation. We find that the parameters $r_1$ and $r_2$ lie within the range [0.9, 1.1], the parameter $q$ remains within the range [-0.05, 0.05], and $\phi$ is within the range [-0.15, 0.15]. This is because the affine transform matrix $T$ is calculated between two consecutive frames. To fairly compare our method with the quantization method proposed in~\cite{Makar:Interframe14}, we also assign 42 bits to the matrix $T$. For $r_1$ and $r_2$, we use 6-bit quantizers, and for $q$, we assign 7 bits. We observed in our experiment that the quantization of $\phi$ strongly affects the matching performance; therefore, 9 bits are assigned to this parameter. For $t_x$ and $t_y$, we assign the same number of bits as in~\cite{Makar:Interframe14} (namely, 7) to enable the exclusive comparison of the quantization methods for the matrix $A$ in~\eqref{eq:affineA}. Subsequently, to improve the matching performance, we increase the allotted space for encoding the matrix $T$ to 48 bits, i.e., 7, 7, 7, 9, 9, and 9 bits for $r_1$, $r_2$, $q$, $\phi$, $t_x$, and $t_y$, respectively, using uniform quantization. The quantized $\overline{T}$ can be obtained from the quantized parameters in~\eqref{eq:affineA}. We use a detection interval of $\Delta$=5 in our experiments for illustration, and the results are shown in Table~\ref{table:Tquantization}. As can be seen, we achieve slightly better results than the method in~\cite{Makar:Interframe14}. Our result for 48 bits approaches the performance achieved using the unquantized $T$. In the following experiments, we use the quantized $\overline{T}$ in place of the uncompressed $T$, using 48 bits to encode the affine transform matrix.
\begin{table}[!htb]
\caption{Comparison of different quantization methods for the affine transform matrix $T$}
\label{table:Tquantization}
\centering 
 
\begin{tabular}{|c|c|}
\hline 
Quantization & \multicolumn{1}{c|}{Avg. \# matches}\tabularnewline
\hline 
\hline 
Uncompressed T & 59.19 (Table~\ref{table:keypointAdaptation}, $\Delta$=5)\tabularnewline
\hline 
Method of \cite{Makar:Interframe14} (42 bits) & 59.00\tabularnewline
\hline 
Our method (42 bits) & 59.06\tabularnewline
\hline 
Our method (48 bits) & 59.13\tabularnewline
\hline 

\end{tabular}

\end{table}

\subsection{Adaptive detection interval} \label{subsec:adaptive}
In the previous section, we presented experiments performed using a fixed detection interval $\Delta$. However, a fixed detection interval will not be adequate when objects are entering or leaving a scene because the descriptors extracted from certain of the estimated keypoints will become spurious. Additionally, when the object of interest does not change across a large number of frames, a fixed detection interval may result in a new set of conventionally detected keypoints being sent sooner than is necessary. To address these issues, we propose to use an adaptive detection interval. Specifically, we will insert a D-frame or a U-frame when the keypoints from the previous D- or U-frame are insufficient for feature extraction in the current frame.

% In this case, we achieve almost a six-fold reduction of keypoint side information compared with that with all D-frames. 

\subsubsection{Adding D-frames or U-frames adaptively}
The proposed process for determining the frame type is as follows. For a new frame, we first extract the descriptors for the estimated keypoints and compare them with the features from the previous D- or U-frame. 

\noindent $\bullet$ If the affine transform matrix cannot be calculated or is incorrect, this indicates that a new scene has begun because the estimated keypoints cannot produce correct descriptors. Therefore, we specify a D-frame for keypoint encoding. 

\noindent $\bullet$ If the number of matches is greater than a certain threshold (e.g., $\epsilon$ = 80\%) with respect to the number of features in the previous D- or U-frame, then most of the estimated keypoints can be considered effective for the current frame. Therefore, an S-frame is specified. Note that the threshold $\epsilon$ significantly affects the determination of S-frames and the bitrate of the side information. For the test dataset, the selected value is reasonable; however, it can be adjusted for other datasets or applications. 

\noindent $\bullet$ If several of the keypoints are still valid, although the object of interest is leaving the current scene, another object is entering the scene, or many keypoints deviate from the actual keypoints after a large number of frames, then we designate the current frame as a U-frame. 

The reason that we compare the features of the current frame with the features of the previous D- or U-frame is that the initial keypoints propagated to the intervening frames originate from this previous D- or U-frame. After determining the frame type, we calculate the affine transform matrix $T$ between the previous frame and the current frame if the frame is designated as an S- or U-frame. Note that we do not use the affine transform matrix between the current frame and the previous D- or U-frame because the compression of this matrix requires a higher bitrate. An individual $T$ for each S- or U-frame is then quantized and transmitted as side information. Transmitting $T$ is not required for D-frames.    

\subsubsection{Comparison of adaptive and fixed detection intervals}
\begin{figure}[!htb]
\centering
\includegraphics[width=0.48 \textwidth]{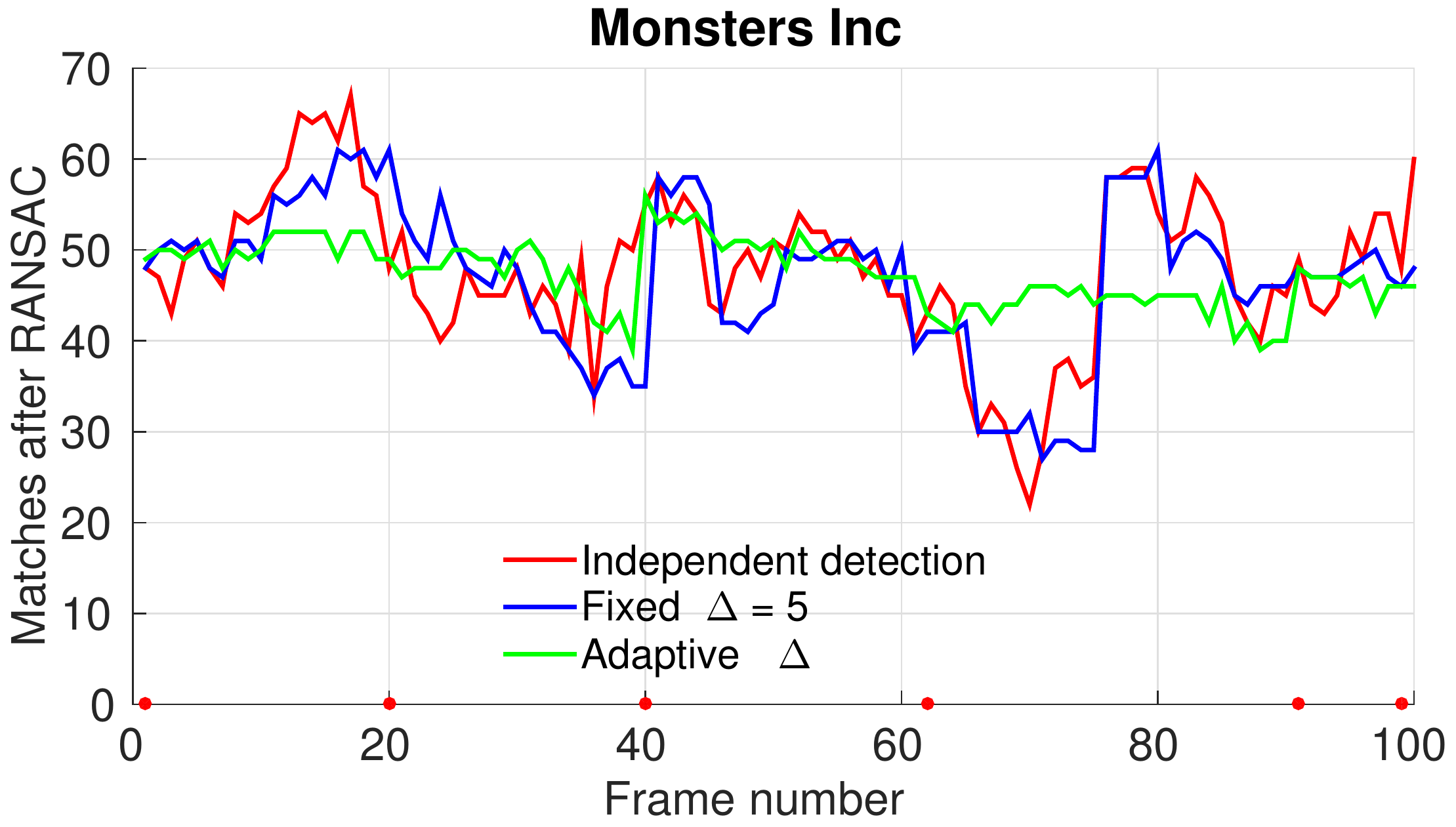} 
%\vspace{-0.1in}
\caption{Matching performance comparison among independent detection, a fixed detection interval ($\Delta$ = 5), and an adaptive detection interval (example video: \textit{Monsters Inc.}). The red dots indicate the frames at which conventional detection is applied in the adaptive-interval scheme.}
\label{fig:compare_adaptive_fixed_monsters}
\end{figure}

In this section, we compare the adaptive-interval and fixed-interval schemes. In this experiment, the previous settings are used: a fixed $\Delta$ = 5, quantization of the affine transform matrix using 48 bits, and conventional feature detection for D- or U-frames. Note that we still use the original (unquantized) keypoints for the D- and U-frames because the objective here is to evaluate the performance of our adaptive detection interval scheme for uncompressed video frames. In addition, we add an \textit{independent detection} scheme to detect features for each frame individually in a conventional manner. From Fig.~\ref{fig:compare_adaptive_fixed_monsters}, we can see that the performance of the fixed-interval scheme is similar to that of the independent detection scheme because the interval $\Delta$ is quite short. The green curve represents the performance of the adaptive-interval scheme, and the red dots represent the D- and U-frames. Interestingly, the keypoints from the D-frames strongly affect the descriptor calculation and matching in the consecutive S-frames. As a result, the adaptive-interval scheme sometimes outperforms the independent detection scheme (e.g., at frame 70), but it is inferior, for example, at frame 80.
% A possible reason is that some of the predicted keypoints produce better descriptors than those calculated using the conventional feature detection process. 
\vspace{-0.1in}

Table~\ref{table:compare_adaptive_fixed} shows the average number of matches and the average number of D- or U-frames over all videos. The fixed-interval scheme achieves similar matching performance to that of the independent detection scheme. In the dataset we are using, each video contains only one object and no scene changes. Therefore, the performance of the fixed-interval scheme does not suffer. The performance of the adaptive-interval scheme is somewhat reduced because of the existence of more S-frames; however, the number of D- and U-frames is significantly reduced. This will result in a significant reduction in the amount of the side information, as shown in Section~\ref{subsec:bitrate_comparison}. Moreover, because of the manner in which we insert D- or U-frames, our adaptive-detection-interval scheme is still suitable for a rapidly changing scene. We will discuss experiments using videos containing scene changes in Section~\ref{subsec:retrievalResults}.
   
\begin{table}[!htb]
\caption{The average numbers of matches and of D- and U-frames over all videos.}
\label{table:compare_adaptive_fixed}
\centering 
 
\begin{tabular}{|c|c|c|}
\hline 
Scheme & Avg. \# matches & Avg. \# D/U-frames\tabularnewline
\hline 
\hline 
Indep. detect. & 59.13 (Table~\ref{table:keypointAdaptation}) & 100\tabularnewline
\hline 
Fixed interval ($\Delta$=5) & 59.13 (Table~\ref{table:Tquantization}) & 20\tabularnewline
\hline 
Adap. interval ($\epsilon$ = 80\%) & 57.95 & 4.63\tabularnewline
\hline 
\end{tabular}
 
\end{table}

\subsection{Switching off keypoint encoding and transmission} \label{subsec:switchoff}
When the camera or one or more objects is moving quickly, the number of features matched in the next frame could fall below the threshold $\epsilon$ = 80\%. In this case, we need to designate a U-frame. Thus, the bitrate of the keypoints will be significantly increased by the necessity of specifying many U-frames for rapidly moving scenes. This situation can be mitigated by reducing the threshold $\epsilon$; however, because this threshold is used to indicate the percentage of well-estimated keypoints, its value should not be set too low. To reduce the bitrate of the side information, as shown in Fig.~\ref{fig:frameTypes}, we add N-frames for quick scene changes. This is motivated by the observation that the video frames for a fast-moving scene are less relevant to the computer vision algorithm run at the server.

The method for determining an N-frame is as follows. Once we have a D- or U-frame, we assume that a new scene or an updated scene is coming. However, if the scene is too short, we may need to add many D- or U-frames, which would introduce a significant increase in bitrate. Therefore, we need to check the length of the scene corresponding to the current D- or U-frame. If the subsequent $N_{s}$ frames are not S-frames, then the current D- or U-frame is changed to an N-frame. This indicates that the scene corresponding to the current frame does not remain stable across $N_{s}$+1 frames. We skip the keypoint encoding for this frame, and the next frame is assumed to be a new D-frame. This process is then performed again for the new D-frame. We switch off the keypoint encoding and transmission for such N-frames, and at the server, the features for these frames are directly extracted from the decoded frames. We will discuss the N-frames selected in a retrieval experiment using \textit{Multiple Objects} video sequences in Section~\ref{subsec:retrievalResults}.

\section{Keypoint encoding and transmission}
\label{sec:keypointencoding}
In the previous sections, we presented several experiments to justify the feasibility of using predicted keypoints, quantizing the affine transform matrix, and using an adaptive detection interval. In addition, we proposed switching off the keypoint encoding and transmission when the scene is moving quickly to reduce the number of D- or U-frames. As a result, the number of keypoints that must be encoded is significantly reduced. Next, we will describe the keypoint encoding approaches used for the different types of frames. First, we use 2 bits to indicate the frame type for each frame. For D-frames, similar to the approach used in our previous study~\cite{Chao:keypointImage} of keypoint encoding for still images, we quantize the keypoint locations into integer values and use sum-based context-based arithmetic coding to encode them. We use 12 bits for scale and orientation encoding. In our experiments, this keypoint encoding method is referred to as the \textit{Intra mode}. For S-frames, we use the estimated keypoints directly and send only the quantized affine transform matrix, which requires 48 bits. This procedure is denoted by \textit{Skip mode} in Fig.~\ref{fig:KeypointEncoderVideo}. For U-frames, we use three modes to encode the keypoints: the \textit{Intra}, \textit{Skip} and \textit{Inter} modes. The Inter mode is identical to the differential keypoint encoding mode shown in Fig.~\ref{fig:KeypointEncoderVideo}. If this mode is selected, then the differences between the scales, locations, and orientations of a matched pair are encoded and transmitted. Note that for an N-frame, the encoding and transmission of the keypoints and the affine transform matrix $T$ are skipped. 

\subsection{Intra mode keypoint encoding} \label{subsec:intraMode}
\subsubsection{Keypoint quantization}
Because the locations, scales, and orientations of the keypoints are represented using floating-point numbers, they must be quantized for efficient transmission. We apply different quantization methods for each.

For the locations, one previously proposed method~\cite{Tsai:improved} uses a spatial grid laid over the original image with step sizes of 4, 6, and 8 pixels. Thus, the locations are quantized by a factor of 4, 6, and 8, respectively. In another study~\cite{Yue:Cloud}, the locations were quantized into integer values, meaning that they were quantized by a factor of 1. In general, locations are quantized as follows:
\begin{equation}
\tilde{l_i} = f \cdot round(\frac{l_i}{f} ) 
\end{equation}
where $f$ is the quantization factor. In our experiments, we determine which quantization factor should be used such that the quantization of the locations does not significantly affect the number of feature matches. Based on the literature~\cite{Chao:keypointImage}, we set $f$ to 1 for location quantization. 

In SIFT, the scale $\sigma_i$ can be represented as follows:
\begin{equation}
\sigma_i = \sigma_0 2^{(o+s/3)} + \Delta\sigma  
\end{equation}
where $\sigma_0$ is a base scale offset (i.e., 2.0159). $o$ is an octave ranging from 0 to a number $x$ that depends on the size of the image ($x$ is less than 4 for our dataset), and $s$ is an integral scale in the range [0, 2]. Thus, 3 and 2 bits are sufficient to represent $o$ and $s$, respectively. $\Delta\sigma$ is an offset calculated to increase the accuracy of the scale estimates in SIFT keypoint detection. In this process, a quadratic polynomial is fit to the values of the detected scale-space extremum ($\sigma_0 2^{(o+s/3)}$) to localize more accurate scales with a resolution that is higher than the scale sampling density. Thus, the value $\Delta\sigma$ is related to the scale-space extremum ($\sigma_0 2^{(o+s/3)}$). We calculate the difference between the detected scale $\sigma_i$ and its corresponding scale-space extremum ($\sigma_0 2^{(o+s/3)}$). We then normalize the difference as follows:
\begin{equation}
\Delta\sigma_n = (\sigma_i - \sigma_0 2^{(o+s/3)}) / \sigma_0 2^{(o+s/3)}  
\end{equation}
Following our previous study~\cite{Chao:keypointImage}, we encode the normalized difference using the Lloyd-Max quantization algorithm. We assign 1 bit to $\Delta\sigma_n$. Thus, the scales, including $o$, $s$, and $\Delta\sigma_n$, are assigned 6 bits in total.    
   
Similar to the location quantization, the orientations are quantized as follows:  
\begin{equation*} 
E(\theta_i) = round( (\frac{\theta_i}{2\pi} + 0.75) \times (2^{t}-1)) \\
\end{equation*}
\begin{equation}\label{eq:index_orientation}
\tilde{\theta_i} = \Big ( \frac{ E(\theta_i) }{ 2^{t}-1 } - 0.75 \Big ) \cdot 2\pi
\end{equation}
where $0.75$ is an offset that ensures that the values $(\frac{\theta_i}{2\pi} + 0.75)$ lie on the interval [0, 1) in the Vlfeat SIFT implementation. Then, the index $E(\theta_i)$ can be represented by values on the interval [0, $2^t$) and can be encoded via fixed-length coding with $t$ bits. In our experiment, $t$ is set to 6, thereby quantizing the orientations into 64 levels.

With this approach, the quantized keypoints yield feature-matching performance similar to that of the ideal, uncompressed keypoints. In total, the encoding of the scale and orientation for one keypoint requires 12 bits.

\subsubsection{Context model for location coding}
As explained in the previous section, we quantize the original locations into integer values (quantization factor 1). To encode the quantized locations, we use the same approach used in CDVS~\cite{MPEG:CDVS}, i.e., sum-based context-based arithmetic coding. The number of bits required for location encoding depends on the number of keypoints to be encoded and their distribution. In sum-based context-based arithmetic coding, we must first train the context model. Similar to the approaches used in a previous study~\cite{Tsai:improved} and in CDVS~\cite{MPEG:CDVS}, we use the INRIA Holidays Dataset\footnote{http://lear.inrialpes.fr/people/jegou/data.php} and the Caltech Building Dataset\footnote{http://vision.caltech.edu/malaa/datasets/caltech-buildings/} for training. Note that the joint dataset comprises 1741 images. Because the location quantization factor is set to 1, the block width is also correspondingly set to 1. Based on the results of our previous study~\cite{Chao:keypointImage}, we select a context range of 49, and the quantized locations are encoded using the corresponding trained model.
 
\subsubsection{Keypoint encoding}
We encode the keypoints using the previously described levels of quantization for their locations, scales, and orientations. If all encoded keypoints $\tilde{k_i}$ are sent as side information, then the extracted descriptors can be expressed as follows:
\begin{equation} \label{eq:encode_decode}
\tilde{d_i} = \Psi( \tilde{k_i} | \overline{I} ), \ with \ \tilde{k_i} = Dec(Enc(k_i)) 
\end{equation}
where $Enc(\cdot)$ and $Dec(\cdot)$ denote the encoding and decoding, respectively, of the keypoints.

\subsection{Keypoint encoding for U-frames} \label{subsec:pmode}
\label{sec:keypointUframe}
For U-frames, we use three modes to encode the keypoints: the \textit{Intra}, \textit{Skip} and \textit{Inter} modes. The Inter mode is the differential keypoint encoding mode shown in Fig.~\ref{fig:KeypointEncoderVideo}. The differences between the scales, locations and orientations of a matched pair are encoded and transmitted. These steps are described as follows. 
\begin{framed}
Step 1. Find the matches between two consecutive frames using the NNDR matching strategy. We denote the matched keypoints in the previous frame by $\mathbf{K^a} = \{ k^a_1, ..., k^a_i, ..., k^a_N \}$ and the matched keypoints in the current frame by $\mathbf{K^b} = \{ k^b_1, ..., k^b_i, ..., k^b_N \}$.

Step 2. Use the quantized affine transform matrix to estimate the locations~\eqref{eq:updateLocation}, scales~\eqref{eq:updateScale}, and orientations~\eqref{eq:updateOrientation} of the keypoints $\mathbf{\hat{K}^b}$ in the current frame from the keypoints $\mathbf{K^a}$ in the previous frame. We denote the keypoint estimation by $\mathbf{K^a} \mapsto \mathbf{\hat{K}^b} = \{ \hat{k}^b_1, ..., \hat{k}^b_i, ..., \hat{k}^b_N \}$. 

Step 3. Calculate the differences to obtain $\Delta l_i = l^b_i-\hat{l}^b_i$, $\Delta \sigma_i = \sigma^b_i - \hat{\sigma}^b_i$ and $\Delta \theta_i = \theta^b_i-\hat{\theta}^b_i$ for each pair $\{ k^b_i, \hat{k}^b_i \}$ of $\{ \mathbf{K^b}, \mathbf{\hat{K}^b} \}$. 

Step 4. Remove incorrect matches if the differences are too large, i.e., $\Delta l_i > 16$, $abs(\Delta \sigma_i) / \hat{\sigma}^b_i > 0.3$, or $abs( Q(\Delta \theta_i) ) > 4$, where $Q(\cdot)$ denotes the index difference derived from~\eqref{eq:index_orientation}. As a result, correct matches have the following properties: $\Delta l_i$ lies on the interval [-16, 16]; the index of $\Delta \sigma_i / \hat{\sigma}^b_i$ lies on the interval [0, 5], where it is quantized into five levels; and $Q(\Delta \theta_i)$ lies on the interval [-4, 4]. 

Step 5: a) The Skip mode is used if the matched keypoints satisfy the following three conditions: 1) $\Delta l_i <= 1$; 2) the index of $\Delta \sigma_i / \hat{\sigma}^b_i$ is equal to 2, which means that the scale has not changed; and 3) $Q(\Delta \theta_i) = 0$. 

b) The Inter mode is used for the differential coding of any other correctly matched keypoints. The differential values are encoded via arithmetic coding. 

c) The Intra mode is employed for non-matched features and incorrectly matched features, which are treated as features corresponding to new scene content. These keypoints are encoded in the same manner as the keypoints in D-frames. 
\end{framed}
\noindent Note that the bitrate for keypoint encoding is determined by the threshold $\epsilon$ that is used to determine the frame types and by the quantization factors that are used for each quantity in the Intra and Inter modes.

\subsection{Bitrate comparison} \label{subsec:bitrate_comparison}
\begin{figure}[!htb]
\centering
\includegraphics[width=0.35 \textwidth]{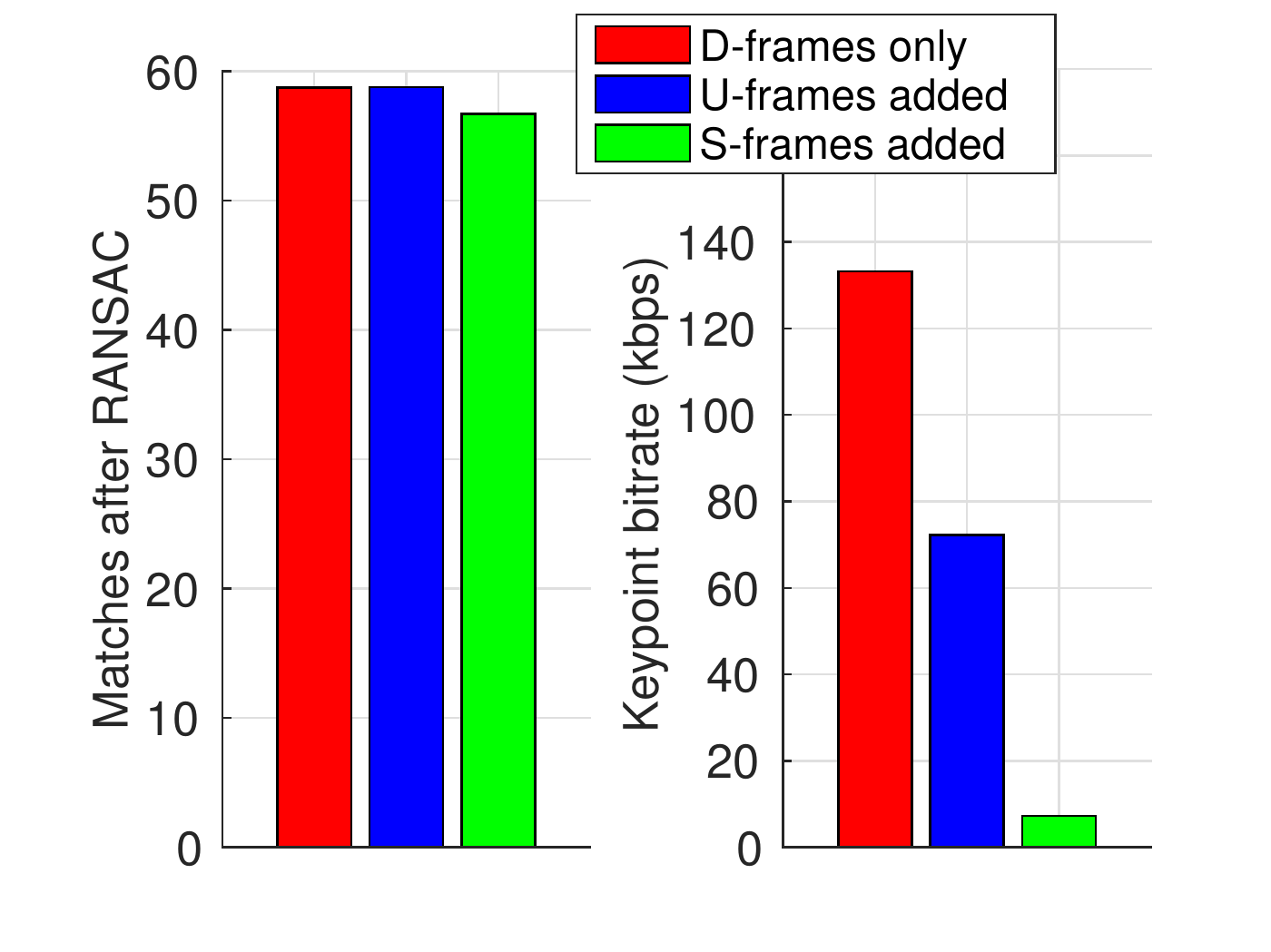} 
\vspace{-0.1in}
\caption{Left: matching performance comparison for the \textit{D-frames only} scheme, the \textit{U-frames added} scheme and the \textit{S-frames added} scheme over all videos. Right: bitrate of the side information.}  
%\vspace{-0.1in}
\label{fig:compare_Matches_bitrate_FrameTypes}
\end{figure}

We encode the keypoints using three different schemes and compare the resulting bitrates. First, all frames are treated as D-frames, which means that all features are independently extracted using the decoded keypoints. Second, only the first frame is considered to be a D-frame, with the following frames being U-frames. Third, S-frames and an adaptive detection interval are added. We compare these schemes to demonstrate the potential bitrate reduction offered by the proposed approach. Fig.~\ref{fig:compare_Matches_bitrate_FrameTypes} shows the results for the three different schemes. The introduction of S-frames and an adaptive detection interval leads to a significant bitrate reduction (by a factor of 18 compared with D-frames only), but the matching performance is not significantly affected.  
 
\begin{figure}[!htb]
\centering
\includegraphics[width=0.48 \textwidth]{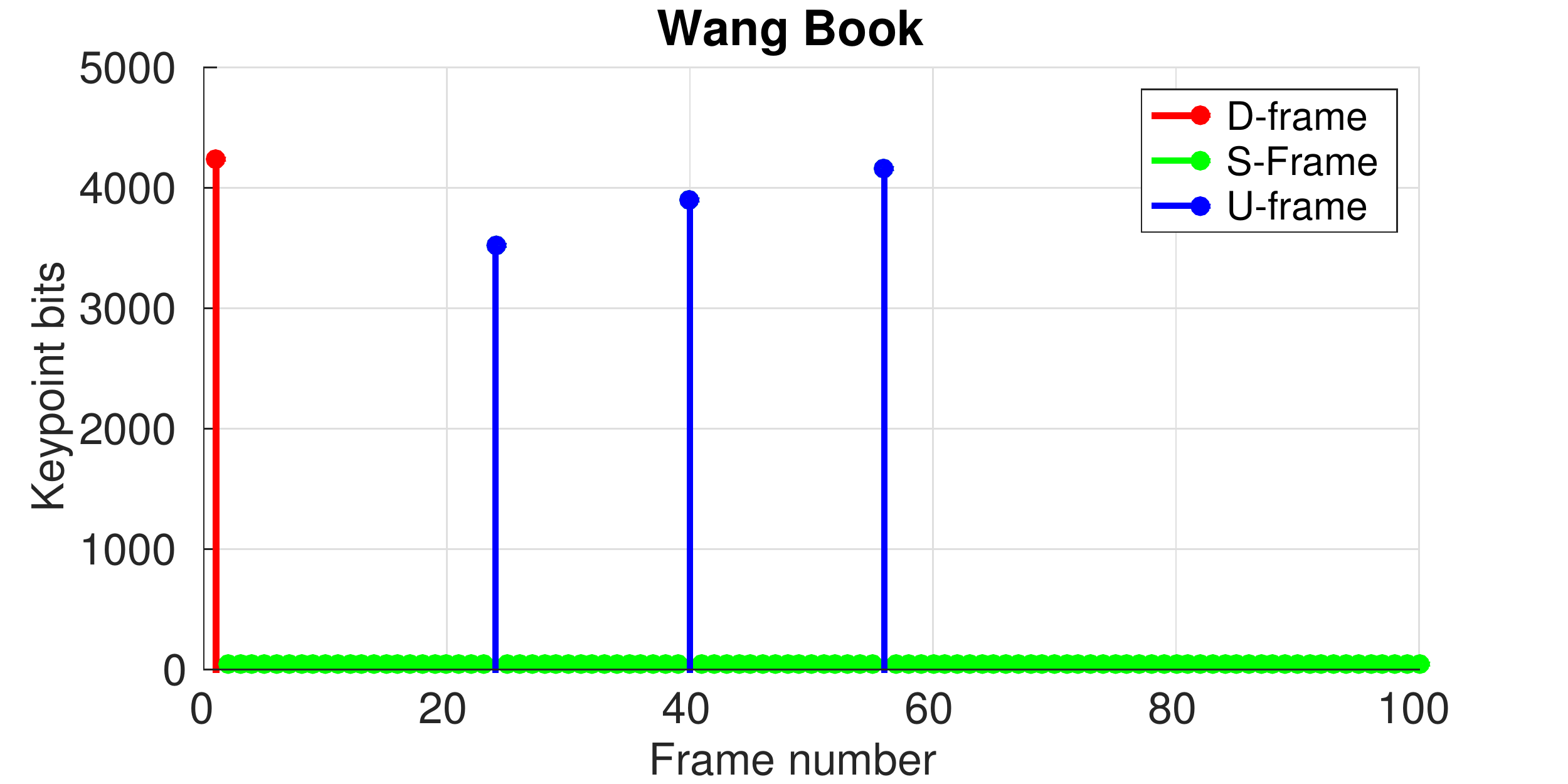} 
\vspace{-0.1in}
\caption{Number of keypoint bits used for each frame (example video: \textit{Wang Book}).}
\label{fig:compare_bits_frameTypes_wangbook}
\vspace{-0.1in}
\end{figure}

Fig.~\ref{fig:compare_bits_frameTypes_wangbook} shows the number of keypoint bits for each frame of the video sequence \textit{Wang Book}. The adaptive detection interval reduces the number of D- or U-frames by exploiting the keypoints that remain coherent across frames. For an S-frame, we use 48 bits for the affine transform matrix and 2 bits to indicate the frame type. For U-frames, many keypoints are encoded in the Intra or Inter mode, therefore requiring a large bitrate. However, incorrect keypoints or deviated keypoints are corrected by adding U-frames.  

%\vspace{-0.1in}

\section{Experimental results}
In the previous section, we described how to encode the keypoints for the different types of frames. Thus far, the encoded keypoints have been used to extract SIFT descriptors from the original video frames. Here, we add the encoded keypoints as side information for compressed videos that are encoded with different QP values. The pairwise matching and retrieval performances are compared with those for standard H.265/HEVC-encoded videos.

\subsection{Pairwise matching results for videos with different QP values}
\begin{figure}[!htb]
\centering
\includegraphics[width=0.48 \textwidth]{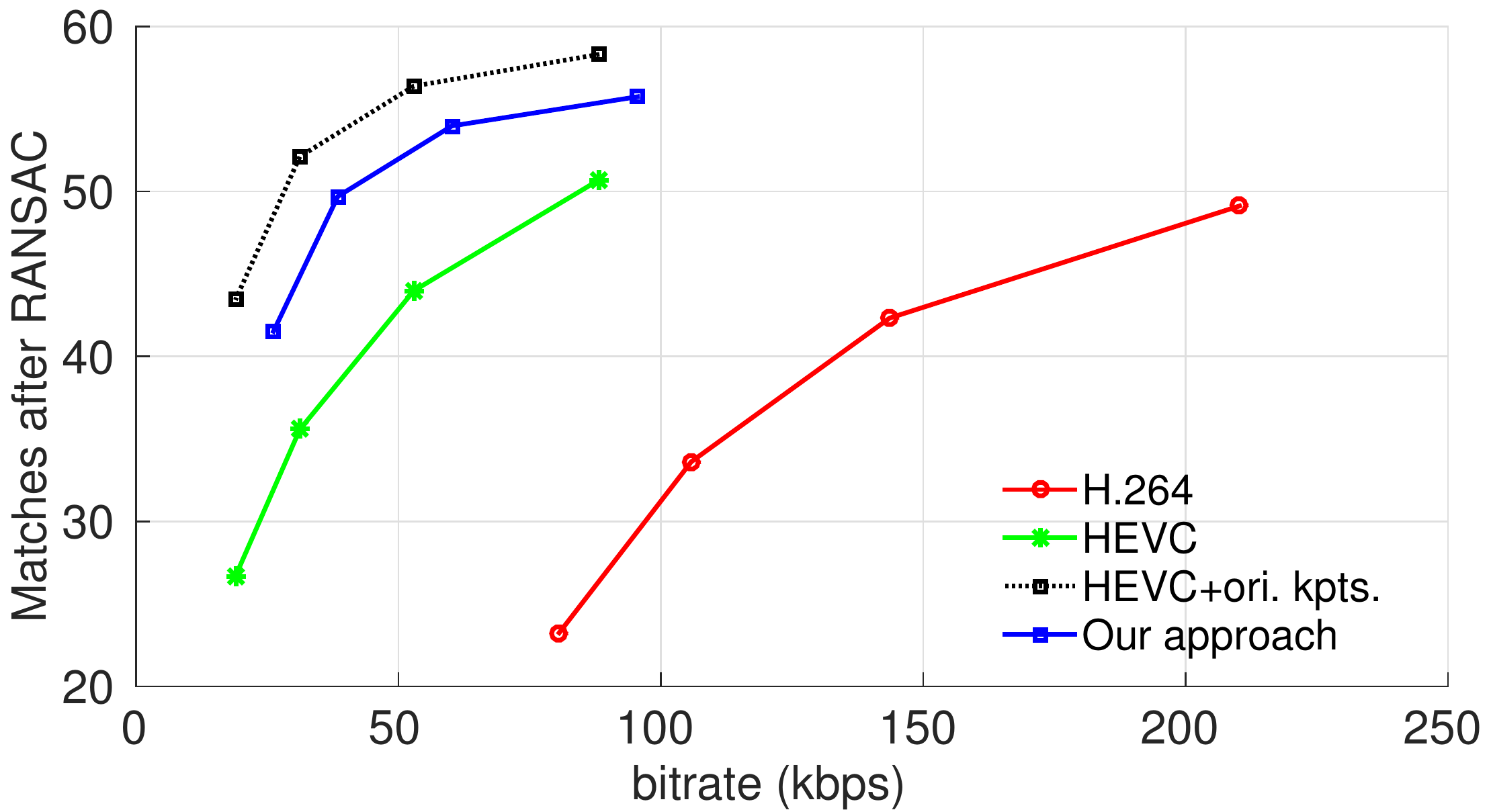} 
\vspace{-0.1in}
\caption{Matching performance comparison for various approaches (the blue line represents the proposed approach). The bitrate includes both the encoded keypoints and the H.265/HEVC-encoded videos.} 
\label{fig:Video_H264_HEVC_Ours}
\end{figure}

First, we perform pairwise matching for the eight videos presented in Section~\ref{exprimentalSettings}. The blue line in Fig.~\ref{fig:Video_H264_HEVC_Ours} shows the final matching performance of the proposed approach as a function of the total bitrate used for the compressed video plus the keypoint side information. The number of matches for a given bitrate budget is significantly improved. In general, a 5$\times$ bitrate reduction is achieved compared with conventional H.264/AVC encoding. This result is better than those for the patch-encoding approaches presented in~\cite{Makar:Interframe} (2.5$\times$) and~\cite{Makar:Interframe14} (4$\times$). Note that in the cited studies, the bitrate reduction was also calculated in comparison with the performance of H.264/AVC encoding; therefore, the comparison is quantitatively fair. Unlike these patch-encoding approaches, we provide standard-compatible videos in addition to the locations, scales, and orientations of the keypoints for a geometric consistency check. The keypoints of a correctly matched image pair should have correlated locations, scales and orientations. Thus, the keypoint information is valuable for eliminating outliers and increasing precision.

\begin{figure}[!htb]
\centering
\vspace{-0.2in}
\includegraphics[width=0.3 \textwidth]{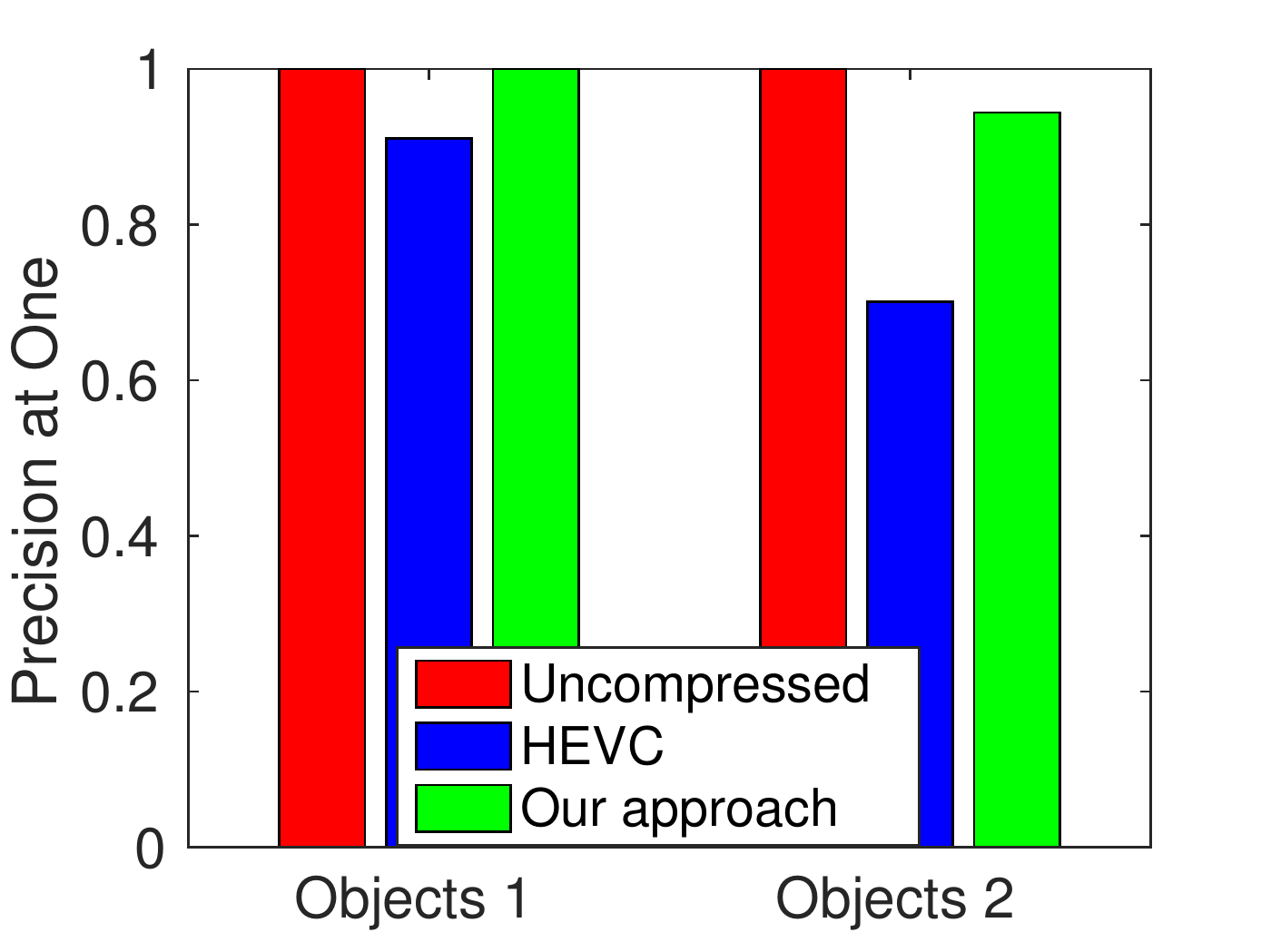} 
% \vspace{-0.1in}
\caption{PAO values for two \textit{Multiple objects} video sequences.} 
\vspace{-0.1in}
\label{fig:retrieval_PAO}
\end{figure}

%%%%%%%%%%%%%%%%%%% note: the following figures were shifted for correctly showing the figure indices.
\begin{figure*}[!htb]
\centering
\includegraphics[width= 0.90 \textwidth]{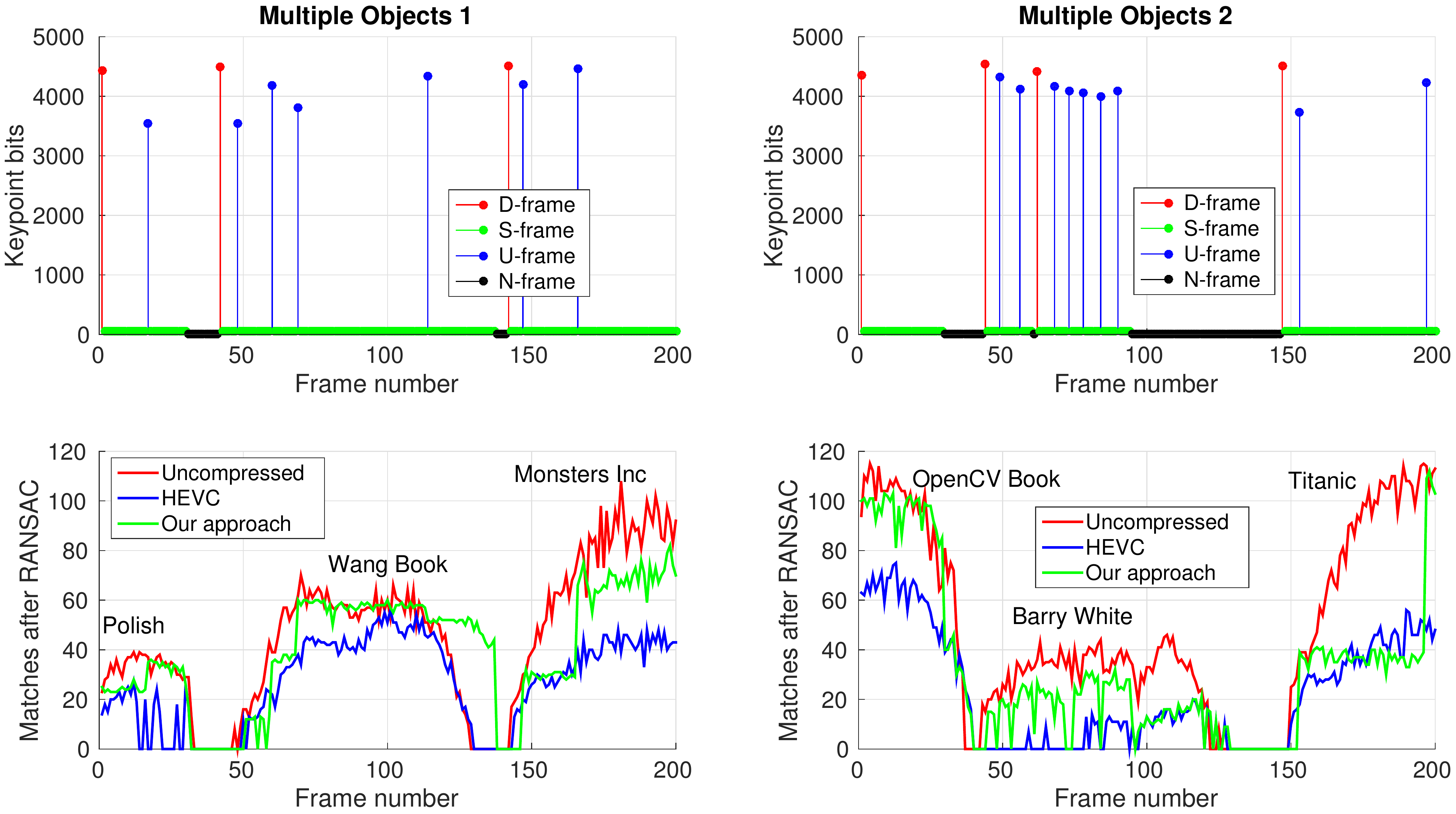}
\vspace{-0.1in}
\caption{The upper plots show the keypoint bits for individual video frames. The lower plots show the pairwise matching performance between the video frame and the top retrieved image.} 
\vspace{-0.1in}
\label{fig:retrieval_all}
\end{figure*}

\subsection{Retrieval results} \label{subsec:retrievalResults}
In our previous experiments, we compared the number of preserved features using pairwise matching. Notably, in content-based image retrieval systems, performance typically improves with an increasing number of preserved features. To evaluate the retrieval performance of our proposed scheme, we use video sequences in the Stanford MAR dataset that show multiple objects. Each \textit{Multiple Objects} video consists of 200 frames and contains three different objects of interest. The first two \textit{Multiple Objects} videos are used in our retrieval experiments. Excluding the fast spatial matching component, we use a previously proposed image retrieval system~\cite{Philbin:AKM}. We use the MIRFLICKR-25000~\cite{Huiskes:MIR} database and the 23 reference images from the Stanford MAR dataset as the training dataset. Similar to~\cite{Makar:Interframe}, we extract up to 300 SIFT descriptors for each image in the database and train one million visual words (VWs) from these descriptors. For the test frames, we extract 200 SIFT features and pass them to the retrieval engine. After obtaining a shortlist of candidate matching images from the retrieval system, we run RANSAC on the top 100 images in the shortlist to reorder these retrieved images for improved precision. We run the retrieval for each frame of the \textit{Multiple Objects} videos. As noted in a previous study~\cite{Makar:Interframe14}, this operation is redundant because the retrieval results for consecutive frames are closely related. However, the objective of this experiment is to examine the performance of different approaches in a scenario wherein an object of interest is leaving or entering the scene. Note that in this experiment, the threshold $\epsilon$ is set to 80\% (Section~\ref{subsec:adaptive}), N-frames are used, and $N_{s}$ (Section~\ref{subsec:switchoff}) is set to 4 for these video sequences. We encode the videos using H.265/HEVC. Note that we use only a QP value of 46 (Section~\ref{subsec:matchesStandards}). Three approaches are compared, i.e., feature extraction from the uncompressed video frames, feature extraction from the compressed video frames, and feature extraction using the encoded keypoints. 

\subsubsection{Precision at One}
Similar to~\cite{Makar:Interframe14}, we first plot the Precision at One (PAO). The PAO is defined as the ratio between the number of correctly retrieved images in the top position and the total number of frames used for retrieval. Note that not all 200 frames are used to calculate the PAO. Only the frames whose locations are tagged in the ground-truth files are used. Fig.~\ref{fig:retrieval_PAO} shows the PAO values for the three tested approaches for the \textit{Multiple Object} videos. The proposed approach based on encoded keypoints yields a significant improvement in terms of the PAO. The bitrates of the two encoded video sequences are 37.08 kbps and 23.44 kbps, and the bitrates for the encoded keypoints are 7.55 kbps and 9.11 kbps, respectively. As seen from the figure, adding the encoded keypoints as side information significantly improves the retrieval performance. In our experiment, we find that the proposed approach offers superior performance compared with videos encoded using a smaller QP value at the same bitrate. 
 
\subsubsection{Number of matches in top retrieved images}
The upper plots of Fig.~\ref{fig:retrieval_all} show the frame types and numbers of bits for individual video frames. The red, green, blue, and black dots represent the keypoint bits for D-, S-, U-, and N-frames, respectively. The keypoint bits for D- and U-frames vary for different video frames, whereas each S-frame requires 50 bits and each N-frame requires 2 bits, thereby resulting in a large bitrate reduction. The lower plots of Fig.~\ref{fig:retrieval_all} show the number of matches achieved using the pairwise matching scheme only when the system retrieves the correct image in the top position. A value of zero indicates that the retrieved image is not in the top position or that no matching image is identified for the video frame. In general, the proposed approach yields an increased number of matches. In the lower right plot, because of the significant amount of glare on the \textit{Barry White} CD cover, many spurious features are extracted across consecutive video frames. These keypoints cannot be propagated to consecutive video frames. Therefore, many frames are selected as N-frames, for which keypoint encoding and transmission are skipped (e.g., video frames 95 to 125), resulting in impaired matching performance for \textit{Barry White}. The results reported in~\cite{Makar:Interframe14} show similar performance impairments. In addition, the glare on the CD cover causes the selection of a greater number of D- or U-frames (e.g., video frames 49 to 90) when encoding the keypoints. There is a drop between frame 155 and frame 195 compared with the uncompressed video, which occurs because the frames during this portion of the video are all of the S-frame type. In Fig.~\ref{fig:compare_adaptive_fixed_monsters}, the number of matches in the current S-frame is closely related to the number of matches in the previous D- or U-frame. From the drop observed in the figure, we can see that the same is also true here. The performance in this portion of the video could be improved by adding a U-frame (i.e., tuning the parameters).

\begin{figure}[!htb]
\centering
\includegraphics[width=0.24 \textwidth]{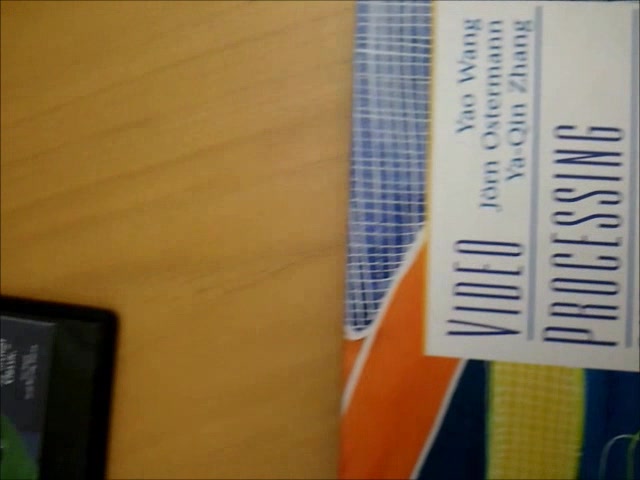} 
\includegraphics[width=0.24 \textwidth]{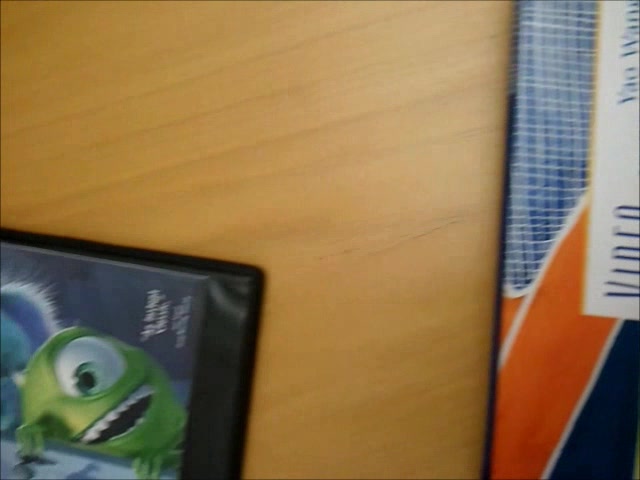} 
%\vspace{-0.1in}
\caption{Transition frames from \textit{Wang Book} to \textit{Monsters Inc.}. Left: frame 135, where the proposed approach correctly detects \textit{Wang Book} and the other two approaches fail. Right: frame 145, where the proposed approach fails to detect the relevant image and the other two approaches successfully detect \textit{Monsters Inc.}.}  
%\vspace{-0.1in}
\label{fig:twoRelevantImages}
\end{figure}

The top detected images for the three approaches differ for certain video frames. For example, as shown in the lower left plot in Fig.~\ref{fig:retrieval_all}, the proposed approach detects \textit{Wang Book} correctly from frame 130 to frame 137, whereas the other two approaches detect no relevant images. By contrast, the other two approaches detect \textit{Monsters Inc.} from frame 144 to frame 146, but the proposed approach fails to detect any relevant image. Fig.~\ref{fig:twoRelevantImages} presents two example frames to illustrate the results. Note that these frames are not included in the calculation of the PAO because the transition sections of the videos are not included in the ground-truth files in the Stanford MAR dataset. 

\subsubsection{Retrieval results for 720p video sequences}
\begin{figure}[!htb]
\centering
\includegraphics[width=0.35 \textwidth]{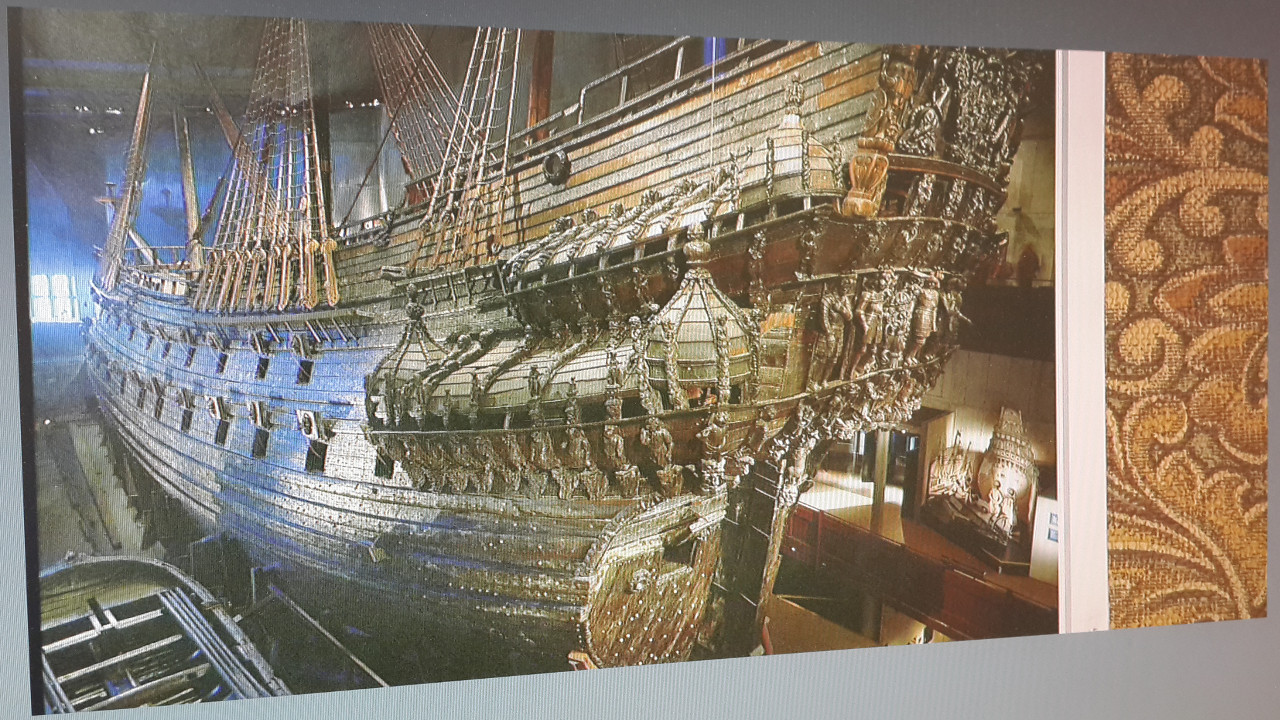} 
\includegraphics[width=0.35 \textwidth]{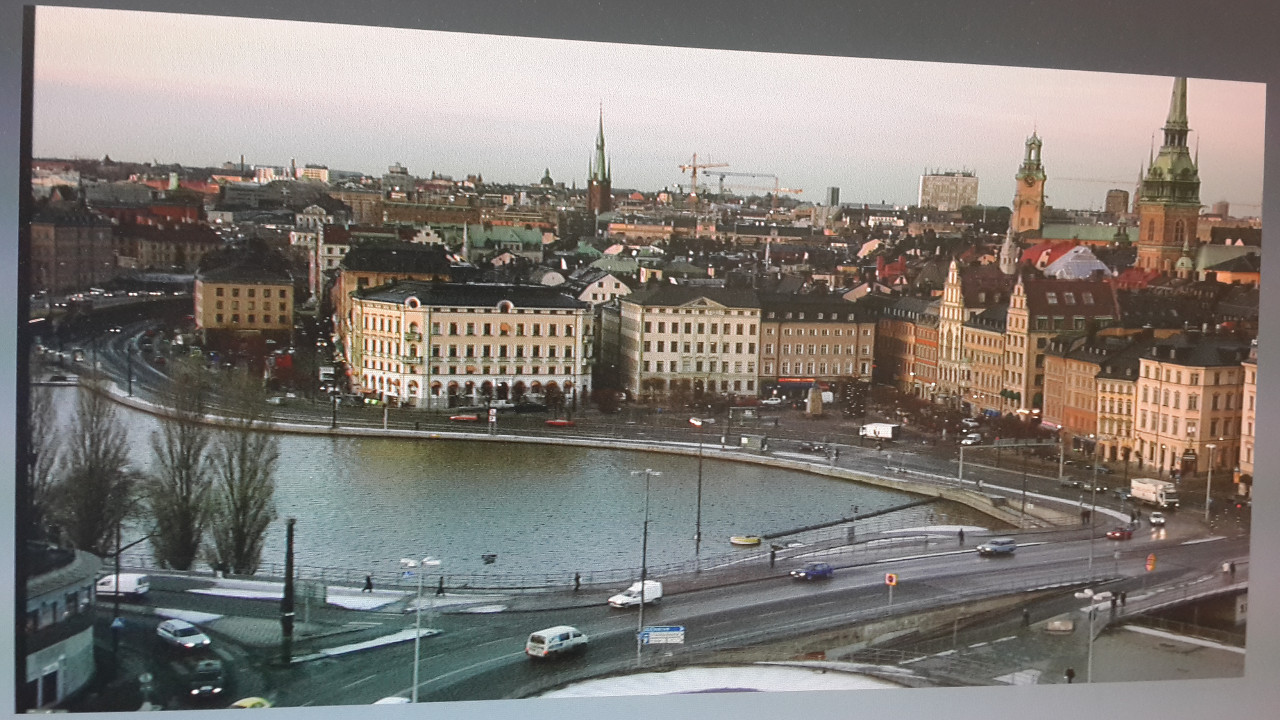} 
\caption{Reference images derived from the first frames of each 720p video.}  
% \vspace{-0.1in}
\label{fig:reference720pImages}
\end{figure}

The Stanford MAR dataset used in the previous experiments is quite small and simple. Therefore, in this retrieval experiment, we use two 720p format video sequences~\cite{TUM:720p} that contain rich features to evaluate our proposed approach: \textit{720p50\_mobcal\_ter} and \textit{720p5994\_stockholm\_ter}. Note that four frames of each video are gray frames, which are removed before coding. We displayed the first frame of each video on a monitor and acquired an image of it using a mobile phone. The resulting images therefore deviate considerably from the original as a result of noise, perspective transformation, illumination changes, and so on. The processed images, shown in Fig.~\ref{fig:reference720pImages}, are used as reference images and integrated into the database used in the previous retrieval experiment to form the training database. 

\begin{figure*}[!htb]
\centering
\includegraphics[width= 0.8 \textwidth]{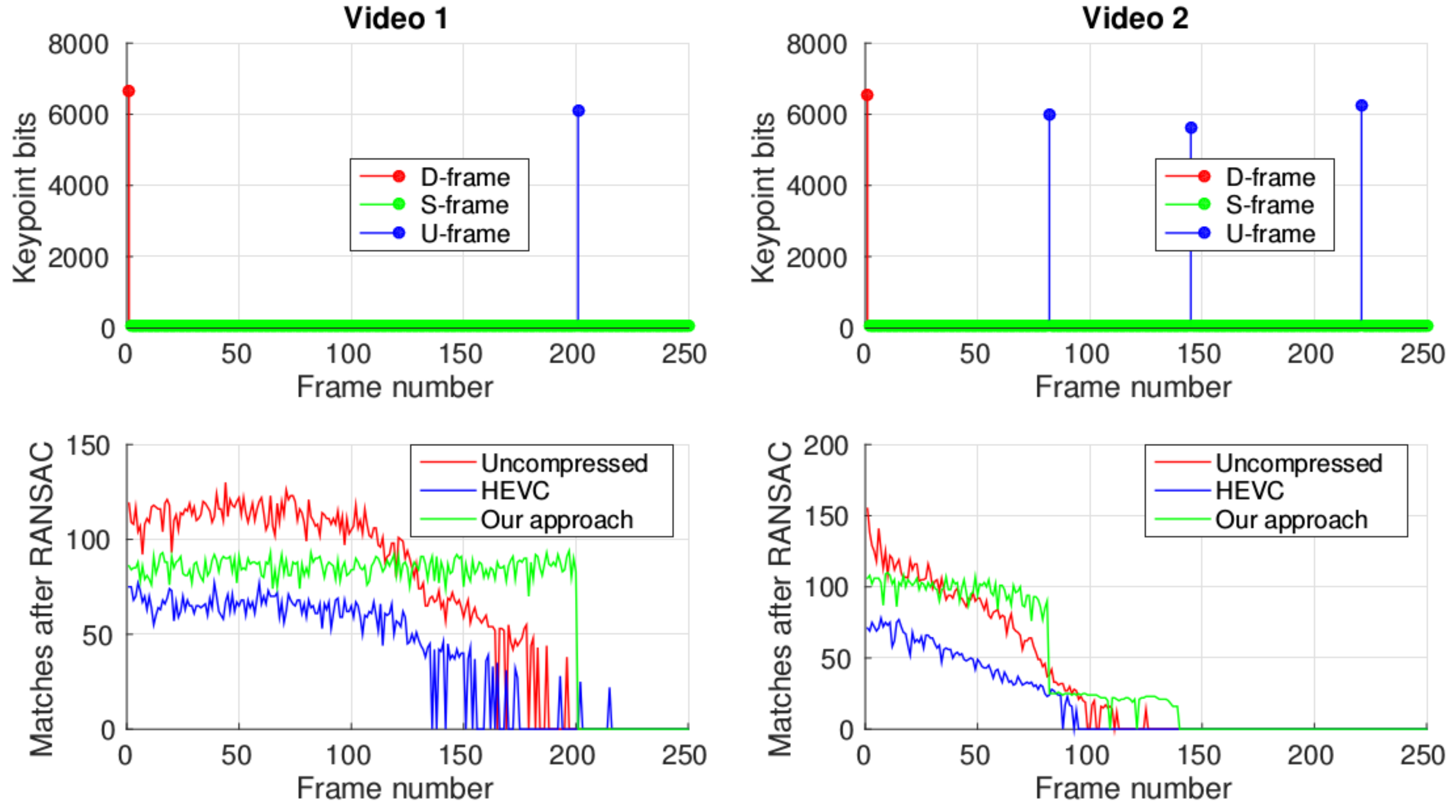}
\vspace{-0.1in}
\caption{The upper and lower plots show the keypoint bits for individual video frames and the pairwise matching performance between the video frame and the top retrieved image, respectively.} 
\vspace{-0.1in}
\label{fig:retrieval_all_720p}
\end{figure*}

\begin{figure}[!htb]
\centering
\includegraphics[width=0.35 \textwidth]{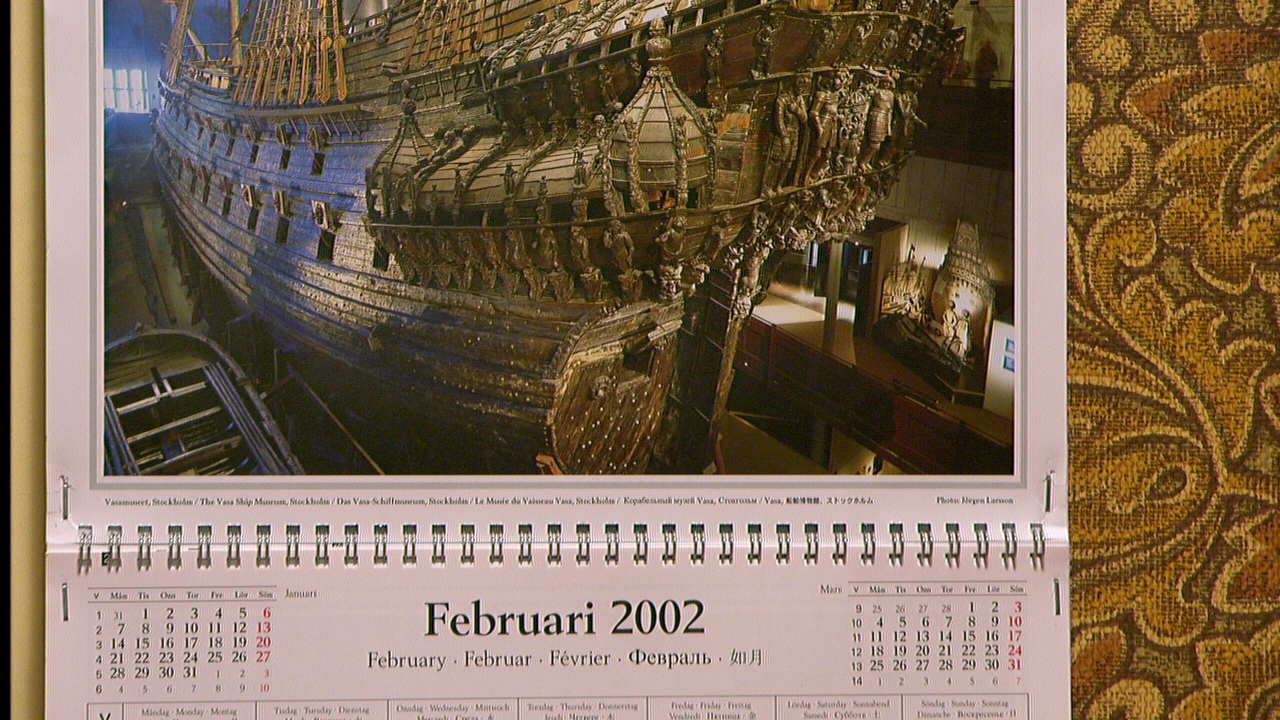} 
\includegraphics[width=0.35 \textwidth]{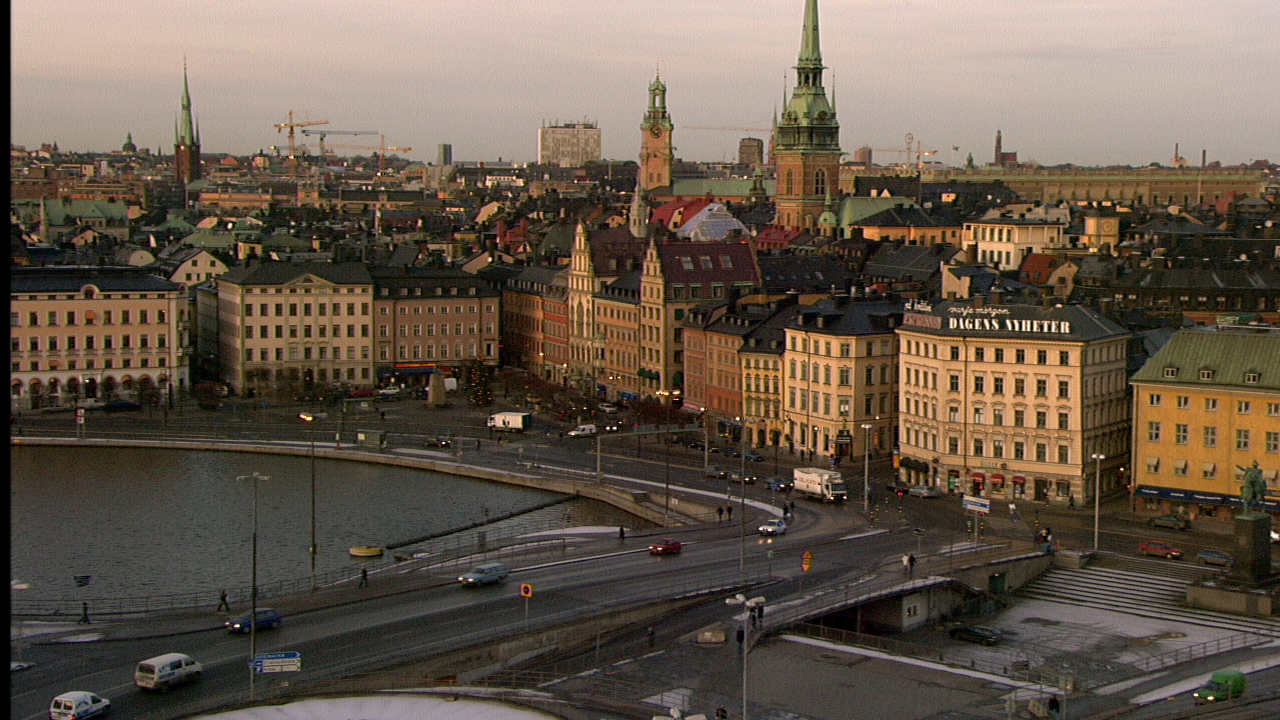} 
%\vspace{-0.1in}
\caption{Examples of retrieved transition frames. Top: frame 200 of \textit{720p50\_mobcal\_ter}. Bottom: frame 139 of \textit{720p5994\_stockholm\_ter}}  
%\vspace{-0.1in}
\label{fig:twoRelevantImages720p}
\end{figure}

As in the previous experiment, the 720p format videos are encoded using H.265/HEVC with a QP value of 46. The keypoints are extracted from the uncompressed video and encoded as side information. Because of the relatively large size of 720p videos, we extract 300 SIFT features from each frame and pass them to the retrieval engine used previously. The other settings are the same as those used in the previous retrieval experiment. The bitrates of the two encoded video sequences are 64.50 kbps and 41.74 kbps, and the bitrates of the encoded keypoints are 9.20 kbps and 6.16 kbps, respectively. We plot the number of matches between the correctly retrieved image (i.e., the reference image is in the top position) and the video frames for the uncompressed videos, the HEVC-encoded videos and the videos encoded using our proposed approach in Fig.~\ref{fig:retrieval_all_720p} (only the results for the first 250 frames are shown). It is apparent that using the encoded keypoints significantly improves the retrieval performance. Note that the content of the first frame gradually disappears in the following frames, and therefore, the number of matches should decrease through consecutive frames. However, in Fig.~\ref{fig:retrieval_all_720p}, the number of matches obtained using our scheme increases significantly. As noted before, the number of matches in subsequent S-frames strongly relies on the number of matches in the previous D-frame, as shown in the figure. In addition, we can see that additional frames can be correctly retrieved with the addition of side information, e.g., frames 175 to 200 in the plot on the lower left and frames 95 to 139 in the plot on the lower right. Compared with the results of the previous retrieval experiment, it is more clearly demonstrated here that the predicted keypoints yield correct descriptors calculated based on these transition frames. It should be noted that the matches identified by our proposed scheme significantly outnumber the matches for the uncompressed video. This can be explained as follows. Only 300 features are extracted from each 720p video frame; therefore, they are highly sparse. Then, a few even stronger features are detected from the coming scene, which are identified as the top 300 features.
%%%Editor - Please ensure that the intended meaning has been maintained in the
%%%above edit.
The previous scene is still present in the video content; thus, the predicted keypoints can yield more valid descriptors than the scheme using the uncompressed video because the first frame is the reference image. Considering the resolution of 720p videos, the scenario depicted in Fig.~\ref{fig:retrieval_all_720p} can rapidly change if more features are detected, i.e., if the number of features is sufficient or if they are sufficiently distributed across the image. Note that no N-frames are specified among the first 250 frames. Fig.~\ref{fig:twoRelevantImages720p} shows two video frames for which only our approach returned a correct match in the top position (see the lower plots in Fig.~\ref{fig:retrieval_all_720p}). We can see that the contents of the query and reference images (see Fig.~\ref{fig:reference720pImages}) overlap to a large extent. Note that the subsequent video frames still contain many features that match with the relevant reference image; however, the relevant reference frame is not returned in the top position. Therefore, the numbers of matches are not shown in Fig.~\ref{fig:retrieval_all_720p}. 

\subsection{Discussion}
\subsubsection{Bitrate reduction for keypoints}
We obtained our results using heuristically selected parameters based on the statistics for keypoint encoding. In the Intra mode procedure presented in Section~\ref{subsec:intraMode}, we quantize the locations using a factor of 1 and use 12 bits to encode the scales and orientations. Note that we can modify these parameters to achieve a larger bitrate reduction for D- and U-frames. In the Inter mode procedure presented in Section~\ref{subsec:pmode}, the differential values of locations, scales, and orientations can be quantized using larger quantizers to further reduce the bitrate for U-frames. The threshold $\epsilon$ (Section~\ref{subsec:adaptive}) determines whether the current frame is designated as a U-frame. Therefore, this parameter strongly affects the length of a series of S-frames. A lower $\epsilon$ value results in a larger number of S-frames for the current D- or U-frame, which yields a lower bitrate for keypoint encoding. Note that this value should not be too small because overly small values will affect the matching performance. The value of $N_{s}$ (Section~\ref{subsec:switchoff}) is used to check the number of S-frames associated with the current D- or U-frame. If the length of the window is too short, then the current D- or U-frame is changed to an N-frame to eliminate the encoding and transmission of keypoints in rapidly moving scenes. Furthermore, in our experiments, it was determined that the acceptable $N_{s}$ value ranges from 3 to 12, depending on the frame rate. In our previous work~\cite{Chao:keypointImage}, we proposed removing spurious keypoints and duplicated keypoints~\cite{Chao:keypointImage} to further lower the number of keypoints to be sent to the server. Note that a spurious keypoint for the current frame could be a useful keypoint for subsequent frames; therefore, we do not directly apply this approach to videos. However, removing duplicated keypoints with respect to the keypoints extracted from the compressed video frame is still feasible. To summarize, there is always a trade-off between the matching performance and the bitrate for keypoint encoding. Note that we present only the results obtained using the parameters selected in the previous sections, which appear reasonably effective for improving the matching performance. 

\subsubsection{Reusing previous descriptors}
For S-frames or in the Skip mode for U-frames, the SIFT descriptors are calculated again at the server based on the estimated keypoints from the compressed frames in our experiments. The purpose of this recalculation is to verify the correctness of the proposed keypoint estimation approach. However, the descriptor calculation process can be skipped by directly using the previous descriptors at the server, i.e., the descriptors for the S-frames and some descriptors for the U-frames can be reused from the descriptors calculated for previous frames. Thus, the computation time at the server can be reduced. In our experiments, from the pairwise matching results shown in Fig.~\ref{fig:Video_H264_HEVC_Ours} and the retrieval results shown in Fig.~\ref{fig:retrieval_all}, it can be seen that the extraction of descriptors from predicted keypoints remains effective. 

\subsubsection{Uploading frames and encoded keypoints}
In certain applications, when a mobile device captures a video, it is unnecessary to upload all video frames. To address this scenario, \cite{Makar:Interframe14} discusses two schemes, both including a \textit{Retrieval State} and a \textit{Pair-wise State}. In the \textit{Retrieval State}, the descriptors extracted from the frame are used to retrieve a relevant image. In the \textit{Pair-wise State}, the descriptors from the current frame are compared with the previously retrieved image using the pairwise matching approach. The authors suggest that better bitrate reduction can be achieved using \textit{On-Device Tracking}~\cite{Makar:Interframe14}. In our proposed system, we can also perform a process that is similar to \textit{On-Device Tracking}~\cite{Makar:Interframe14}. When a new object is detected, the frame is HEVC Intra-encoded, and the keypoints are encoded using the Intra mode and sent as side information with the encoded frame. Note that this is similar to our previously proposed approach for images~\cite{Chao:keypointImage}, which achieves improved performance at low bitrates. The server then performs image retrieval and sends the relevant image back to the mobile device. Then, the descriptors from the following frames are compared with the retrieved image. If no new object is found, then there is no need for data transmission between the mobile device and the server. 

\subsubsection{Computational complexity}
It should be noted that the processes of keypoint detection, feature extraction and matching, and keypoint encoding are all based on uncompressed video frames. Therefore, in practice, we can run these processes in parallel with the actual video compression, as shown in Fig.~\ref{fig:idea_architecture}, to speed up the processing. In addition, to reduce computational complexity, we can replace the descriptor extraction and descriptor matching processes with simpler detectors, descriptors, and matching procedures, e.g., we can heuristically determine which keypoints can be used in the next frame. This keypoint encoding process can be optimized to achieve a reduced computation time compared with the H.265/HEVC encoding time. Moreover, because the keypoint decoding is separate from the decoding of the video, the encoded bitstreams need not be transmitted simultaneously with the video. In certain applications, they can be transmitted later when the communication channel is not busy to improve the matching performance.    

\section{Conclusion}
\label{sec:conclusion}
Because of the adverse effects of video compression on feature-matching performance, we propose to encode the original SIFT keypoints from a video and transmit them along with the compressed video to the server. In this paper, we introduce four different types of frames for keypoint encoding based on considerations regarding different behaviors in consecutive video frames. Then, we propose methods of predicting keypoints, quantizing the affine transform matrix, adopting an adaptive detection interval, and switching off the keypoint encoding when the scene is moving quickly to reduce the keypoint bitrate. We describe the Intra, Inter and Skip modes of encoding the keypoints. Finally, pairwise matching and image retrieval are performed. The results show that the proposed approach achieves improved performance in feature matching at a given rate. The proposed feature-preserving video compression approach is advantageous because a standard-compatible video can be watched or stored for future use, flexible feature types can be extracted, and the orientations and scales can be used for 
geometric verification. In addition, when more features (e.g., 500 features) must be transmitted, the increase in bitrate incurred by our proposed scheme is much smaller than that of other schemes. Moreover, other types of keypoints (e.g., SURF~\cite{Bay:SURF06}, MSER~\cite{Matas:MSER}, and FAST~\cite{Rosten:FAST}) can be similarly encoded for improved feature extraction using the proposed framework.

% or


%\appendices
%% you can choose not to have a title for an appendix
%% if you want by leaving the argument blank
%\section{}
%% use section* for acknowledgment
%\section*{Acknowledgment}

\ifCLASSOPTIONcaptionsoff
  \newpage
\fi



%\bibliographystyle{IEEEtran}

\bibliographystyle{IEEEtran}
\bibliography{IEEEabrv,IEEEexample}

\end{document}